\documentclass[english]{svjour3}
\usepackage[latin9]{inputenc}
\usepackage{array}
\usepackage{float}
\usepackage{mathtools}
\usepackage{multirow}
\usepackage{amstext}
\usepackage{graphicx}

\makeatletter

\providecommand{\tabularnewline}{\\}
\floatstyle{ruled}
\newfloat{algorithm}{tbp}{loa}
\providecommand{\algorithmname}{Algorithm}
\floatname{algorithm}{\protect\algorithmname}

\usepackage{cite}
\usepackage{url}
\usepackage{graphicx}
\usepackage{setspace}
\usepackage{graphicx}
\usepackage{graphics} 
\usepackage{epsfig} 
\usepackage{mathptmx} 
\usepackage{times} 
\usepackage{amsmath} 
\usepackage{amssymb}  
\usepackage{epstopdf}
\usepackage[final]{changes}
\usepackage{xcolor}
\usepackage{bbm}
\usepackage{tikz}
\usetikzlibrary{arrows}
\usepackage{caption}
\usepackage{algpseudocode}
\usepackage{mathtools}
\usepackage[para]{threeparttable}
\usepackage{color}
\usepackage{mathtools, cuted}

\@ifundefined{showcaptionsetup}{}{%
 \PassOptionsToPackage{caption=false}{subfig}}
\usepackage{subfig}
\makeatother

\usepackage{babel}

\begin{document}
\title{\LARGE \bf Multi-Sound-Source Localization Using Machine Learning for Small Autonomous Unmanned Vehicles with a Self-Rotating Bi-Microphone Array }
\author{Deepak~Gala$^{1}$, Nathan~Lindsay$^{2}$, and Liang Sun$^{3}$
\thanks{$^{1}$Deepak~Gala is with the Klipsch School of Electrical and Computer Engineering, New Mexico State University, Las Cruces, NM, 88001 USA         {\tt\small drgala@nmsu.edu}}%
\thanks{$^{2}$Nathan~Lindsay is with the Department of Mechanical and Aerospace Engineering, New Mexico State University, Las Cruces, NM, 88001 USA         {\tt\small nl22@nmsu.edu}}%
\thanks{$^{3}$Liang Sun is with the Department of Mechanical and Aerospace Engineering, New Mexico State University, Las Cruces, NM, 88001 USA         {\tt\small lsun@nmsu.edu}}%
}

\maketitle
\thispagestyle{empty}
\pagestyle{empty}
\begin{abstract}
While vision-based localization techniques have been widely studied
for small autonomous unmanned vehicles (SAUVs), sound-source localization
capabilities have not been fully enabled for SAUVs. This paper presents
two novel approaches for SAUVs to perform three-dimensional (3D) multi-sound-sources
localization (MSSL) using only the inter-channel time difference (ICTD)
signal generated by a self-rotating bi-microphone array. The proposed
two approaches are based on two machine learning techniques viz.,
Density-Based Spatial Clustering of Applications with Noise (DBSCAN)
and Random Sample Consensus (RANSAC) algorithms, respectively, whose
performances are tested and compared in both simulations and experiments.
The results show that both approaches are capable of correctly identifying
the number of sound sources along with their 3D orientations in a
reverberant environment. 
\end{abstract}

\section{Introduction}

Small autonomous unmanned vehicles (SAUVs, e.g., quadcopters and ground
robots) have revolutionized civilian and military missions by creating
a platform for observation and permitting access to locations that
are too dangerous, too difficult or too costly to send humans. The
sensing capability of SAUVs has been enabled by various sensors, such
as RGB cameras, infrared cameras, LiDARs, RADARs, and ultrasound sensors.
However, these mainstream sensors are subject to either lighting conditions
or line-of-sight requirements. On the other end of the spectrum, as
the sound travels to all directions and can be transmitted even with
some small obstacles~\cite{Wang2016}, sound sensors have great potential
to overcome the line-of-sight constraints of the aforementioned sensors
and provide SAUVs with an omnidirectional full span sensing coverage.
Sound-based sensing capabilities would significantly facilitate SAUVs
in critical applications (e.g., search and rescue) and enable sociable
and service robots (e.g., shopping assistants and restaurant waiters)
to collaborate with humans in complicated scenarios~\cite{Boehme2003,Murray2004}.

Among the sensing tasks for SAUVs, localization is of utmost significance~\cite{Borenstein1996}.
While vision-based localization techniques have been developed based
on cameras, sound source localization (SSL) has been achieved using
microphone arrays with different numbers (e.g., 2, 4, 8, 16) of microphones.
Although it has been reported that the accuracy of the localization
is enhanced as the number of microphones increases~\cite{Rabinkin1998,brandstein2013microphone},
this comes with a price of algorithm complexity and hardware cost,
especially due to the expense of Analog-to-Digital converters (ADC),
which is proportional to the number of speaker channels. 

Humans and many other animals can locate sound sources with decent
accuracy and responsiveness by using their two ears associated with
head rotations to avoid ambiguity (i.e., cone of confusion)~\cite{Wallach1939}.
SSL techniques using a self-rotating bi-microphone array have been
reported in the literature~\cite{Lee2015,Handzel2002,Eriksen2006,Zhong2015,Gala2018}.
To eliminate cone of confusion~\cite{Wallach1939}, the bi-microphone
array is rotated around the center of the robot/dummy head on the
horizontal plane so that an Inter-Channel Time Difference (ICTD) signal
is generated whose data points form multiple discontinuous sinusoidal
waveforms. Single-SSL (SSSL) techniques with different numbers of
microphones have been well-studied~\cite{Valin2003}, while reported
multi-sound-source-localization (MSSL) techniques typically require
large microphone arrays with specific structures, which are not easy
to be mounted on SAUVs. Pioneer work for MSSL assumed the number of
sources to be known beforehand~\cite{Sun2014,zhong2016active}. Some
of these approaches~\cite{Blandin2012,Swartling2011,Dong2013} are
based on sparse component analysis (SCA) that requires the sources
to be W-disjoint orthogonal~\cite{Yilmaz2004} (i.e., in some time-frequency
components, at most one source is active), thereby making them unsuitable
for reverberant environments. Pavlidi et al.~\cite{Pavlidi2013}
and Loesch et al.~\cite{Loesch2008} presented an SCA-based method
that counts and localizes multiple sound sources but requires one
sound source to be dominant over others in a time-frequency zone. 

Clustering methods have also been used to conduct MSSL~\cite{Swartling2011,Blandin2012,Dong2013}.
Catalbas et al.~\cite{Catalbas2017} presented an approach for MSSL
by using four microphones and the sound sources are required to be
present within a predefined boundary. The technique was limited to
localize sound orientations in the two-dimensional plane using k-medoids
clustering. The number of sound sources was calculated using the exhaustive
elbow method, which is instinctive and computationally expensive.
Traa et al.~\cite{Traa2013} presented an approach that utilizes
the time-delay between the microphones in the frequency domain to
model the phase differences in each frequency bin of a short-time
Fourier transform. Using the linear relationship between phase difference
and frequency, the data were then clustered using random sample consensus
(RANSAC). In our previous work~\cite{Gala2018,Zhong2015,Gala2019Moving,Gala2019},
we developed an SSSL technique based on an extended Kalman filter
and an MSSL technique based on a cross-correlation approach, which
was computationally expensive. 

The contributions of this paper include two novel MSSL approaches
for SAUVs. Both approaches are able to identify both the number of
sound sources and their 3D locations by only using a self-rotating
bi-microphone array. In the first approach, a novel mapping mechanism
is developed to convert the acquired ICTD signal to an orientation
domain. Unsupervised classification is then conducted using the Density-Based
Spatial Clustering of Applications with Noise (DBSCAN)~\cite{Ester1996}.
The second approach is based on a sinusoidal ICTD regression using
a RANSAC-based method. Both simulations and experiments were conducted
to verify the proposed methodology. 

The rest of the paper is organized as follows. In Section~\ref{sec:Preliminaries},
the mathematical calculation for the ICTD signal generated by the
self-rotating microphone array is presented. In Section~\ref{sec:Model-for-Mapping},
the mapping mechanism for regression and clustering is presented.
Section~ \ref{sec:Methodology} presents the two proposed approaches
for MSSL. Simulation and experimental results are presented and discussed
in Section~\ref{sec:Results-and-Discussion}. Section~\ref{sec:Conclusion-and-Future-Scope}
concludes the paper.

\section{Preliminaries\label{sec:Preliminaries}}

\subsection{Inter-Channel Time Difference (ICTD)}

The ICTD is the time difference between a sound signal arriving at
two microphones and can be calculated using the cross-correlation
technique~\cite{Knapp1976The,azaria1984time}. Consider a single
stationary sound source and two spatially separated microphones placed
in an environment. Let $y_{1}(t)$ and $y_{2}(t)$ be the sound signals
captured by the microphones in presence of noise, which are given
by~\cite{Knapp1976The} $y_{1}(t)=s(t)+n_{1}(t)$ and $y_{2}(t)=\delta\cdot s\left(t+t_{d}\right)+n_{2}(t)$,
where $s\left(t\right)$ is the sound signal, $n_{1}(t)$ and $n_{2}(t)$
are real and jointly stationary random noises, \emph{$t_{d}$} denotes
the time difference of $s\left(t\right)$ arriving at the two microphones,
and $\delta$ is the signal attenuation factor due to different traveling
distances. It is commonly assumed that $\delta$ changes slowly and
\emph{$s(t)$} is uncorrelated with noises $n_{1}(t)$ and $n_{2}(t)$~\cite{Knapp1976The}. 

The cross-correlation of $y_{1}$ and $y_{2}$ is given by $R_{y_{1},y_{2}}(\tau)=E\left[y_{1}(t)\cdot y_{2}(t-\tau)\right]$,
where \emph{$E\left[\cdot\right]$} represents the expectation operator.
Various pre-filters that eliminate or reduce the effect of background
noise and reverberations have been used prior to the cross-correlation~\cite{naylor2010speech,Gala2010,Gala2011}.
The time difference of $y_{1}$ and $y_{2}$, i.e., the ICTD, is given
by $\hat{T}\triangleq\arg\max_{\tau}R_{y_{1},y_{2}}.$ The distance
difference of the sound signal traveling to the two microphones is
given by $d\triangleq\hat{T}\cdot c_{0},$ where $c_{0}$ is the sound
speed and is usually selected as 345 m/s on the Earth surface. 
\begin{remark}
For simplicity, the signal $d$ is referred as ICTD in this paper.
ICTD is the only cue used in this paper for the source counting and
localization, while no aforementioned scaling functions nor pre-filters
are used.
\end{remark}

\subsection{Far-Field Assumption\label{subsec:Far-Field-Assumption}}

The five different fields around a sound source are free field, near
field, far field, direct field, and reverberant field~\cite{ISO12001,Hansen2001}.
The region where the sound pressure and the acoustic particle velocity
are not in phase is regarded as the near field. The far field of a
source begins where the near field ends and extends to infinity. Under
the far-field assumption, the acoustic wavefront reaching the microphones
is planar and not spherical, in the sense that the waves travel in
parallel. This means that the angle of incidence will be the same
for the two microphones. Further, it can be shown that with $D/b>2.7$,
the error of the far-field approximation drops below $0.5^{o}$, where
$D$ is the distance of the sound source to the center of the microphone
array and $b$ is the distance between the microphones~\cite{Calmes2009}.

\subsection{Mathematical Model for ICTD signal \label{sec:Mathematical-Model}}

In this paper, the location of a single sound source is defined in
a spherical coordinate frame, whose origin is assumed to coincide
with the center of a ground robot. 
\begin{figure}[h]
\centering{}\includegraphics[bb=0bp 0bp 491bp 358bp,width=0.7\columnwidth]{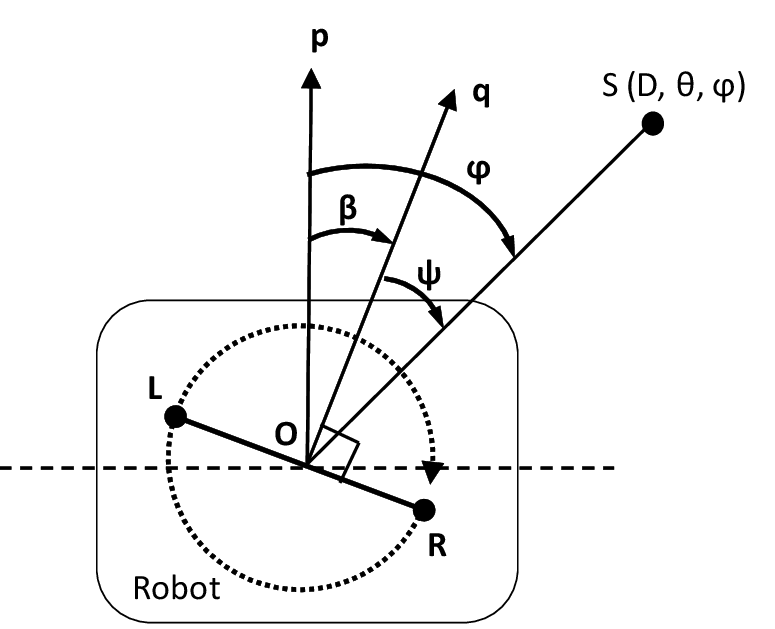}\caption{Top-down view of the system. \label{figure system formulation}}
\end{figure}
\begin{figure}[h]
\centering{}\includegraphics[width=0.7\columnwidth]{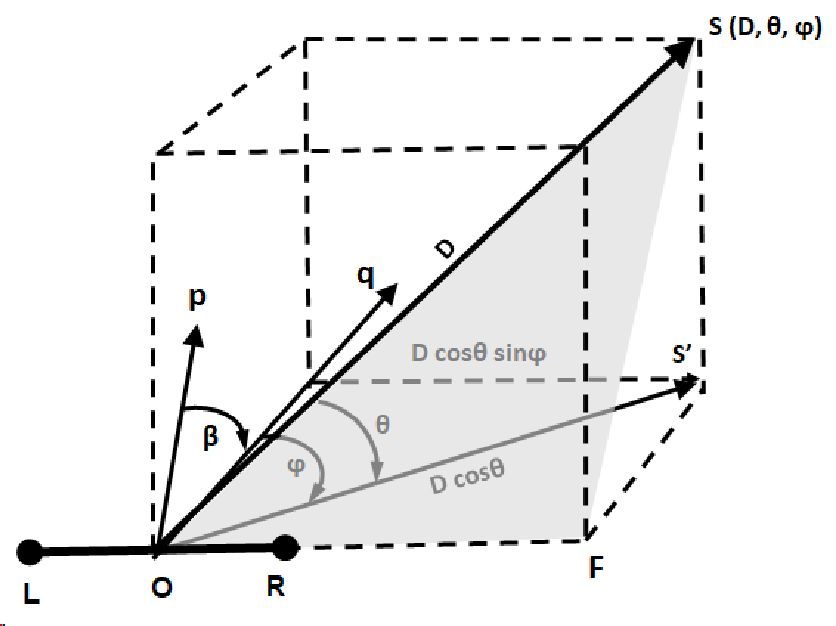}\caption{3D view of the system.\label{fig:Spherical Coordinate System}}
\end{figure}
\begin{figure}[h]
\centering{}\includegraphics[width=0.7\columnwidth]{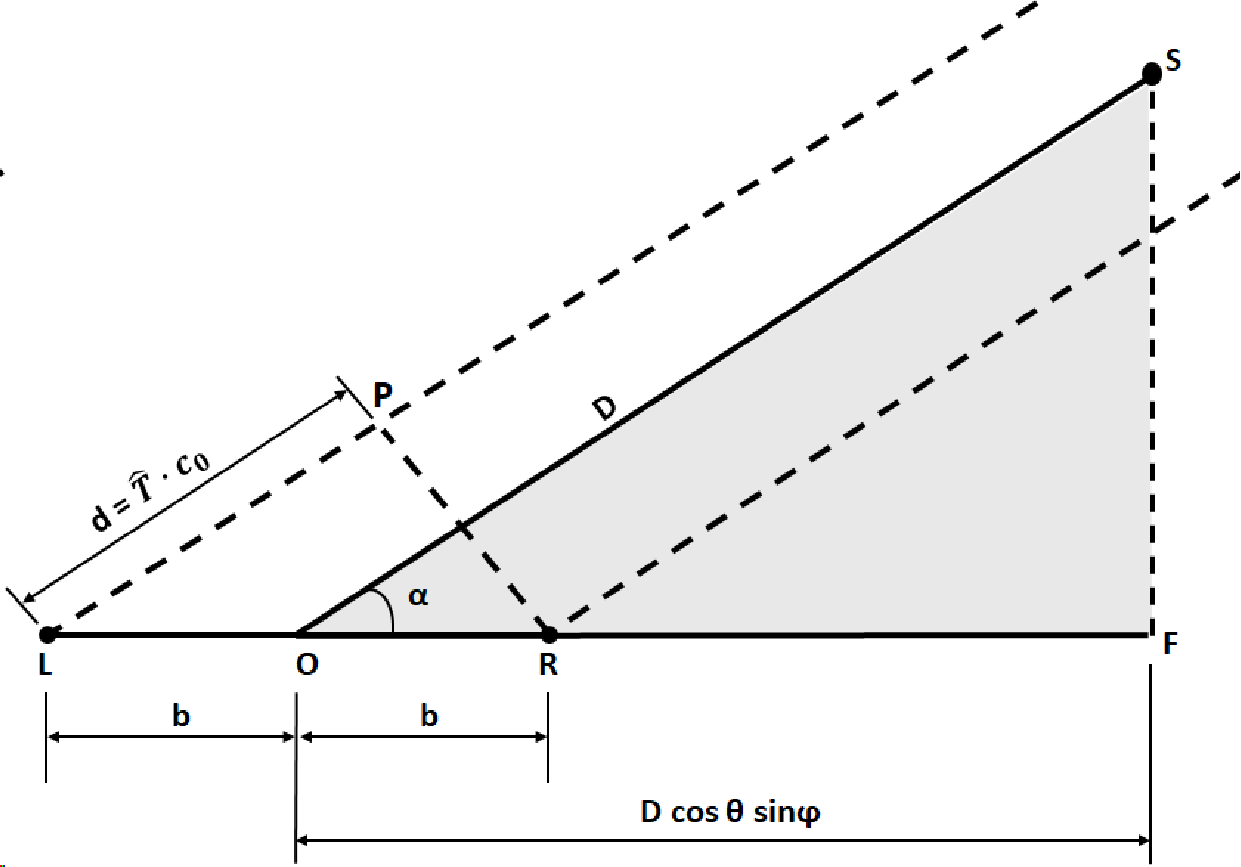}\caption{Top-down view of the plane containing triangle $SOF$.\label{fig:Gray Triangle Plane}}
\end{figure}

As shown in Figs.~\ref{figure system formulation} and \ref{fig:Spherical Coordinate System},
the left and right microphones, $L$ and $R$, collects the acoustic
signal generated by the sound source \emph{S}. Let $O$ be the center
of the robot as well as the bi-microphone array. The sound source
location is represented by ($D,\theta,\varphi$), where $D$ is the
distance between the source and the center of the robot, i.e., the
length of segment $\overline{OS}$, $\theta\in\left[0,\frac{\pi}{2}\right]$
is the elevation angle defined as the angle between $\overline{OS}$
and the horizontal plane, and $\varphi\in(-\pi,\pi${]} is the azimuth
angle defined as the angle measured clockwise from the robot heading
vector, $\mathbf{p}$, to $\overline{OS'}$. Letting unit vector $\mathbf{q}$
be the orientation (heading) of the microphone array, $\beta$ be
the angle from $\mathbf{p}$ to $\mathbf{q}$, and $\psi$ be the
angle from $\mathbf{q}$ to $\overline{OS'}$, both following a clockwise
rotation rule, we have
\begin{equation}
\varphi=\psi+\beta.\label{eq:azimuth}
\end{equation}
In the shaded triangle, $SOF$, shown in Figs\emph{\large{}.~}\ref{fig:Spherical Coordinate System}
and \ref{fig:Gray Triangle Plane}, define $\alpha=\angle SOF$ and
we have $\cos\alpha=\cos\theta\sin\psi.$ Based on the far-field assumption~\cite{ISO12001},
we have
\begin{equation}
d\triangleq\hat{T}\cdot c_{0}=2b\cos\alpha=2b\cos\theta\sin\psi.\label{eq:d}
\end{equation}

To avoid cone of confusion~\cite{Wallach1939} in SSL, we are considering
an ICTD signal generated by a microphone pair with time-varying positions,
such as a self-rotating bi-microphone array. Without loss of generality,
in this paper, we assume a clockwise rotation of the microphone array
on the horizontal plane, while the robot itself does not rotate throughout
the entire estimation process, which implies that $\varphi$ in Equation~\eqref{eq:azimuth}
is constant. The rotation of the microphone array is assumed to be
triggered by a sound detection mechanism, which is beyond the scope
of this paper and hence will not be discussed.

The initial heading of the microphone array is configured to coincide
with the heading of the robot, i.e., $\beta\left(t=0\right)=0$, which
implies that $\varphi=\psi\left(0\right)$. As the microphone array
rotates clockwise with a constant angular velocity, $\omega$, we
have $\beta\left(t\right)=\omega t$ and due to Equation~\eqref{eq:azimuth}
we have $\psi\left(t\right)=\varphi-\beta\left(t\right)=\varphi-\omega t.$
The resulting time-varying $d\left(t\right)$ due to Equation~\eqref{eq:d}
is then given by
\begin{align}
d\left(t\right) & =2b\cos\theta\sin\left(-\omega t+\varphi\right).\label{eq:ITD_equation}
\end{align}
Because the microphone array rotates on the horizontal plane, $\theta$
does not change during the rotation for a stationary sound source.
The resulting $d\left(t\right)$ is a sinusoidal signal with the amplitude
$A\triangleq2b\cos\theta$, which implies that 
\begin{equation}
\theta=\cos^{-1}\frac{A}{2b}.\label{eq:Theta-from-A}
\end{equation}
 It can be seen from Equation~\eqref{eq:ITD_equation} that the phase
angle of $d\left(t\right)$ is the azimuth angle of the sound source.
Therefore, the location (i.e., azimuth and elevation angle) of the
sound source can by determined by estimating the characteristics (i.e.,
the amplitude and phase angle) of the sinusoidal signal, $d(t)$.

The collection of the ICTD signal for multiple sound sources (as shown
in Fig.~\ref{fig:Mult-Source-ITD-Estimation}) illustrates a group
of multiple discontinuous sinusoidal waveforms, each of which corresponds
to a single sound source, satisfying the amplitude-elevation and phase-azimuth
relationship as mentioned above.

\section{Model for Mapping and Sinusoidal Regression\label{sec:Model-for-Mapping}}

The signal $d(t)$ in Equation~\eqref{eq:ITD_equation} is sinusoidal
with its amplitude $A=2b\cos\theta$ and phase angle $\varphi$ that
corresponds to the azimuth angle of the sound source. Since the frequency,
$\omega$, of $d\left(t\right)$ is the known rotational speed of
the microphone array, the localization task (i.e., identifying $\theta$
and $\ensuremath{\varphi}$) is to estimate the amplitude and phase
angle of $d\left(t\right)$, i.e., $A$ and $\beta$. 

Consider a general form of $d(t)$ expressed as 
\begin{equation}
d(t)=A_{1}\text{s}_{\omega t}+A_{2}\text{c}_{\omega t},
\end{equation}
where $\text{s}_{\omega t}=\sin(\omega t)$ and $\text{c}_{\omega t}=\cos(\omega t)$,
and we have $A=\sqrt{A_{1}^{2}+A_{2}^{2}},$ and $\varphi=\tan^{-1}\left(\frac{A_{2}}{A_{1}}\right).$
Consider two data points $y_{1}=d(t_{1})$ and $y_{2}=d(t_{2})$,
collected at two distinct time instants $t_{1}$ and $t_{2}$, respectively,
and we have
\begin{equation}
\left[\begin{array}{c}
y_{1}\\
y_{2}
\end{array}\right]=\left(\begin{array}{cc}
\text{s}_{\omega t_{1}} & \text{c}_{\omega t_{1}}\\
\text{s}_{\omega t_{2}} & \text{c}_{\omega t_{2}}
\end{array}\right)\left[\begin{array}{c}
A_{1}\\
A_{2}
\end{array}\right].
\end{equation}
If $\text{s}_{\omega(t_{2}-t_{1})}\neq0,$ then we can obtain 
\begin{equation}
A=\frac{\sqrt{y_{1}^{2}+y_{2}^{2}-2y_{1}y_{2}\text{c}_{\omega(t_{2}-t_{1})}}}{\text{s}_{\omega(t_{2}-t_{1})}},\label{eq:Amplitude-two-points}
\end{equation}
and 
\begin{equation}
\varphi=\tan^{-1}\left(\frac{y_{1}\text{s}_{\omega t_{2}}-y_{2}\text{s}_{\omega t_{1}}}{y_{2}\text{c}_{\omega t_{1}}-y_{1}\text{c}_{\omega t_{2}}}\right).\label{eq:Azimuth-two-points}
\end{equation}

\section{Methodology\label{sec:Methodology}}

\subsection{DBSCAN-Based MSSL\label{sec:Localization-using-DBSCAN}}

DBSCAN is one of the most popular nonlinear clustering techniques
and it can discover any arbitrarily shaped clusters of densely grouped
points in a data set and outperform other clustering methods in the
literature~\cite{Raj2017,Farmani2017}. In the DBSCAN algorithm~\cite{Raj2017},
a random point from the data set is considered as a core cluster point
when more than $m$ points (including itself) within a distance of
$\epsilon$ (epsilon ball) exists in its neighborhood. This cluster
is then extended by checking all of the other points satisfying the
$\epsilon-m$ criteria thereby letting the cluster grow. A new arbitrary
point is then chosen and the process is repeated. The point which
is not a part of any cluster and having fewer than $m$ points in
its $\epsilon$-ball is considered as a ``noise point''. The DBSCAN
technique is more suitable for applications with noise and outperforms
the $k$-means method, which requires a prior knowledge of the number
and the approximate initial centroids of clusters and is highly sensitive
to noisy data points and to the selection of the initial centroids~\cite{Celebi2013}.

The proposed DBSCAN-based MSSL technique consists of two stages. In
the first stage, the data points of the ICTD signal are mapped to
the orientation (i.e., the elevation-azimuth coordinate) domain. The
data set consisting of all data points in a multi-source ICTD signal
contains not only inliers but also outliers, which produce undesired
mapped locations. When the number of inliers is significantly greater
than the outliers after a number of iterations, highly dense clusters
will be formed. In the second stage, these clusters are detected using
the DBSCAN technique by carefully selecting parameters $m$ and $\epsilon$.
The number of clusters corresponds to the number of sound sources
and the centroids of these clusters represent the locations of the
sound sources.

The complete DBSCAN-based MSSL algorithm is described in Algorithm~\ref{alg:DBSCAN-algo}.
Two points in the data set are selected randomly and mapped into the
orientation domain by calculating angles $\theta$ and $\varphi$
using Equations~\eqref{eq:ITD_equation}, \eqref{eq:Amplitude-two-points},
\eqref{eq:Theta-from-A} and~\eqref{eq:Azimuth-two-points}. A set
$M\coloneqq\{(\theta_{1},\varphi_{1}),$ $(\theta_{2},\varphi_{2}),$
$\text{..., }(\theta_{N},\varphi_{N})\}$ of these mapped points is
then created. The process for detection of clusters is then started.
A point $(\theta_{i},\varphi_{i})$ in $M$ is randomly chosen and
is decided to be a core cluster point or a noise point by checking
the density-reachability criteria under the $m$-$\epsilon$ condition~\cite{Raj2017}.
The time complexity of Algorithm~\ref{alg:DBSCAN-algo} is $\mathcal{O}(N_{D}^{2})$,
where $N_{D}$ is the number of iterations for mapping and clustering.
The selection of parameters for Algorithm~\ref{alg:DBSCAN-algo}
will be discussed in Section~\ref{subsec:Parameter-Selection}.
\begin{algorithm}[h]
1: Capture $d(t)$ for one full rotation of the bi-microphone array

2: Select $m$\textbf{ }and\textbf{ }$\epsilon$

3: Select the number of iterations $N_{D}$

4:\textbf{ FOR} $i=1$ to $N_{D}$ \textbf{DO} 

5: $\text{ \text{ \text{ \text{ }}}}$ $\text{ \text{ \text{ \text{ }}}}$Randomly
choose non-repeated set of two points $y_{1}$ and $y_{2}$ from $d$,
such that $y_{1}\neq y_{2}$ and do not equal zero simultaneously

6:$\text{ \text{ \text{ \text{ }}}}$ $\text{ \text{ \text{ \text{ }}}}$
Calculate $\hat{A}$ and $\hat{\varphi}$ using Equations~\eqref{eq:Amplitude-two-points}
and \eqref{eq:Azimuth-two-points}

7: $\text{ \text{ \text{ \text{ }}}}$$\text{ \text{ \text{ \text{ }}}}$
Calculate $\hat{\theta}_{i}$ using Equation~\eqref{eq:Theta-from-A}
and $\hat{\varphi}_{i}=\hat{\varphi}$

8:\textbf{ END FOR}

9: \textbf{FOR }$i=1$ to $N_{D}$ \textbf{DO}

10: $\text{ \text{ \text{ \text{ }}}}$ $\text{ \text{ \text{ }}}$Randomly
choose the pair $(\theta_{i},\varphi_{i})$ from the set $M\coloneqq\{(\theta_{1},\varphi_{1}),\text{ }(\theta_{2},\varphi_{2})\text{ ..., }(\theta_{N_{D}},\varphi_{N_{D}})\}$

11: $\text{ \text{ \text{ \text{ }}}}$ $\text{ \text{ }}$ Calculate
the distance between the chosen $(\theta_{i},\varphi_{i})$ and every
other point in $S$ 

12: $\text{ \text{ \text{ \text{ }}}}$ $\text{ \text{ \text{ }}}$\textbf{IF
}the number of points in the range $\epsilon$ is greater than $m$

13: $\text{ \text{ \text{ \text{ }}}}$ $\text{ \text{ \text{ \text{ }}}}$$\text{ \text{ \text{ \text{ }}}}$
Label $(\theta_{i},\varphi_{i})$ as a core cluster point

14: $\text{ \text{ \text{ \text{ }}}}$ $\text{ \text{ \text{ }}}$\textbf{ELSE}

15: $\text{ \text{ \text{ \text{ }}}}$ $\text{ \text{ \text{ }}}$$\text{ \text{\text{ }}}$$\text{ \text{\text{ }}}$Label
$(\theta_{i},\varphi_{i})$ as a noise point 

16: $\text{ \text{ \text{ \text{ }}}}$ $\text{ \text{ \text{ }}}$\textbf{END
IF}

17: \textbf{END FOR}

\caption{DBSCAN-Based MSSL\label{alg:DBSCAN-algo}}
\end{algorithm}

\subsection{RANSAC-Based MSSL\label{sec:Localization-using-RANSAC}}

The RANSAC~\cite{Fischler1981} algorithm iteratively uses a set
of observed data to estimate parameters of a mathematical model and
can identify inliers (e.g., parameters of a mathematical model) in
a data set that may contain a significantly large number of outliers.
The input to the RANSAC algorithm includes a set of data, a parameterized
model, and a confidence parameter ($\sigma_{conf}$). In each iteration,
a subset of the original data is randomly selected and used to fit
the predefined parameterized model. All other data points in the original
data set are then tested against the fitted model. A point is determined
to be an inlier of the fitted model, if it satisfies the $\sigma_{conf}$
condition. The process is repeated by selecting another random subset
of the data. After a number of iterations, the parameters are then
selected for the best fitting (with maximum inliers) estimated model.
\begin{algorithm}[h]
1: Capture $d(t)$ for one full rotation of the bi-microphone array

2: Select $N_{R}$, $\sigma_{conf}$ and initialize $e=0$

3: \textbf{WHILE }there are samples in $d$

4:$\text{ \text{ \text{ \text{ }}}}$ $\text{ \text{ \text{ \text{ }}}}$\textbf{FOR}
$j=1$ to $N_{R}$ \textbf{DO} 

5: $\text{ \text{ \text{ \text{ }}}}$ $\text{ \text{ \text{ \text{ }}}}$Randomly
choose non-repeated set of two points $y_{1}$ and $y_{2}$ from $d$,
such that $y_{1}\neq y_{2}$ and do not equal zero simultaneously

6:$\text{ \text{ \text{ \text{ }}}}$$\text{ \text{ \text{ \text{ }}}}$
$\text{ \text{ \text{ \text{ }}}}$ Calculate $\hat{A}$ and $\hat{\varphi}$
using Equations~\eqref{eq:Amplitude-two-points} and~\eqref{eq:Azimuth-two-points}

7: $\text{ \text{ \text{ \text{ }}}}$$\text{ \text{ \text{ \text{ }}}}$$\text{ \text{ \text{ \text{ }}}}$
Calculate $\hat{d}=\hat{A}\sin(\omega t+\hat{\varphi})$

8: $\text{ \text{ \text{ \text{ }}}}$$\text{ \text{ \text{ \text{ }}}}$$\text{ \text{ \text{ \text{ }}}}$
Calculate $count$ = number of points in $d$ fitting $\hat{d}$ with
at least $\sigma_{conf}$

9: $\text{ \text{ \text{ \text{ }}}}$$\text{ \text{ \text{ \text{ }}}}$$\text{ \text{ \text{ \text{ }}}}$
\textbf{IF }$e<count$

10:$\text{ \text{ \text{ \text{ }}}}$ $\text{ \text{ \text{ \text{ }}}}$$\text{ \text{ \text{ \text{ }}}}$$\text{ \text{ \text{ \text{ }}}}$\textbf{$A_{K}=\hat{A}$,
$\varphi_{K}=\hat{\varphi}$ }and\textbf{ $e=count$}

11: $\text{ \text{\text{ \text{ }}}}$ $\text{ \text{ \text{ \text{ }}}}$$\text{ \text{ \text{ }}}$\textbf{END
IF}

12:$\text{ \text{ \text{ \text{ }}}}$$\text{ \text{ \text{ \text{ }}}}$\textbf{END
FOR}

13:$\text{ \text{ \text{ \text{ }}}}$ $\text{ \text{ \text{ }}}$Calculate
$\theta_{K}$ using Equation~\eqref{eq:Theta-from-A}

14:$\text{ \text{ \text{ \text{ }}}}$ $\text{ \text{ \text{ }}}$$e=0$

15:$\text{ \text{ \text{ \text{ }}}}$ $\text{ \text{ \text{ }}}$Remove
samples on $\hat{d}$ within $\sigma_{conf}$ from $d$

16:\textbf{ END WHILE}

\caption{RANSAC-Based MSSL \label{alg:RANSAC-algo}}
\end{algorithm}

The RANSAC-based MSSL method is described in Algorithm~\ref{alg:RANSAC-algo}.
It can be seen from Equation~\eqref{eq:ITD_equation} that the signal
$d(t)$ generated by the self-rotating bi-microphone array is sinusoidal.
Two points from the ICTD signal are selected randomly and a sine wave
with the given frequency (i.e., the angular speed of the rotation,
$\omega)$ is generated. The $count$ represents the number of points
whose distance to the fitted sine wave is less than $\sigma_{conf}$,
which is the threshold for a point to be considered inlier. Then the
points in $d$ that belong to $\hat{d}$ according to the $\sigma_{conf}$
condition will be removed from $d$, This procedure is repeated for
$N_{R}$ iterations and the parameters $A_{K}$ and $B_{K}$ are updated
every time the number of inliers is greater than that in the previous
iterations. This process is repeated until either all the points in
$d$ are examined or $N_{R}$ iterations are completed. The time complexity
of Algorithm~\ref{alg:RANSAC-algo} is $\mathcal{O}(\frac{n}{2}\cdot N_{R})$,
where $n$ is the number of samples in the ICTD. After the first few
of $N_{R}$ iterations, most of the data points are removed. This
results in $n$ to be a small number as compared to $N_{R}$. 

\subsection{Parameter Selection\label{subsec:Parameter-Selection}}

The value of $\sigma_{conf}$ can be chosen depending on the possibility
of sound sources to be close to each other and the noise level. The
maximum number of possible unique combinations of randomly chosen
data points $y_{1}$ and $y_{2}$ is $C(u,v)=\frac{u!}{(u-v)!\text{ }v!}$,
where the value of $u$ is the number of data points and $v$ is the
number of randomly chosen data points. However, the number of iterations,
$N_{D}$, needs only to be large enough for the efficient formation
of the clusters. The larger the value of $N_{D}$, the denser the
clusters would be, which will further modify the $\epsilon$ value
in Algorithm~\ref{alg:DBSCAN-algo}. On the other hand, $N_{R}$
should be chosen large enough to ensure that at least one of the sets
of randomly selected points does not include an outlier. The value
of the $N_{R}$ can be smaller than $N_{D}$ due to the removal of
all the data points within the $\sigma_{conf}$ range, which reduces
the data points to further deal with. Further, these values also depends
on how noisy the data is. The number of iterations for RANSAC-based
MSSL can be calculated using $N_{R}=\frac{\log(1-s)}{\log(1-(1-\epsilon)^{w})}$,
where $s$ is the probability of success, $\epsilon$ is outlier ratio
and $w$ is the number of required points to fit the model. 

\begin{figure}[h]
\hfill{}\includegraphics[width=0.98\columnwidth]{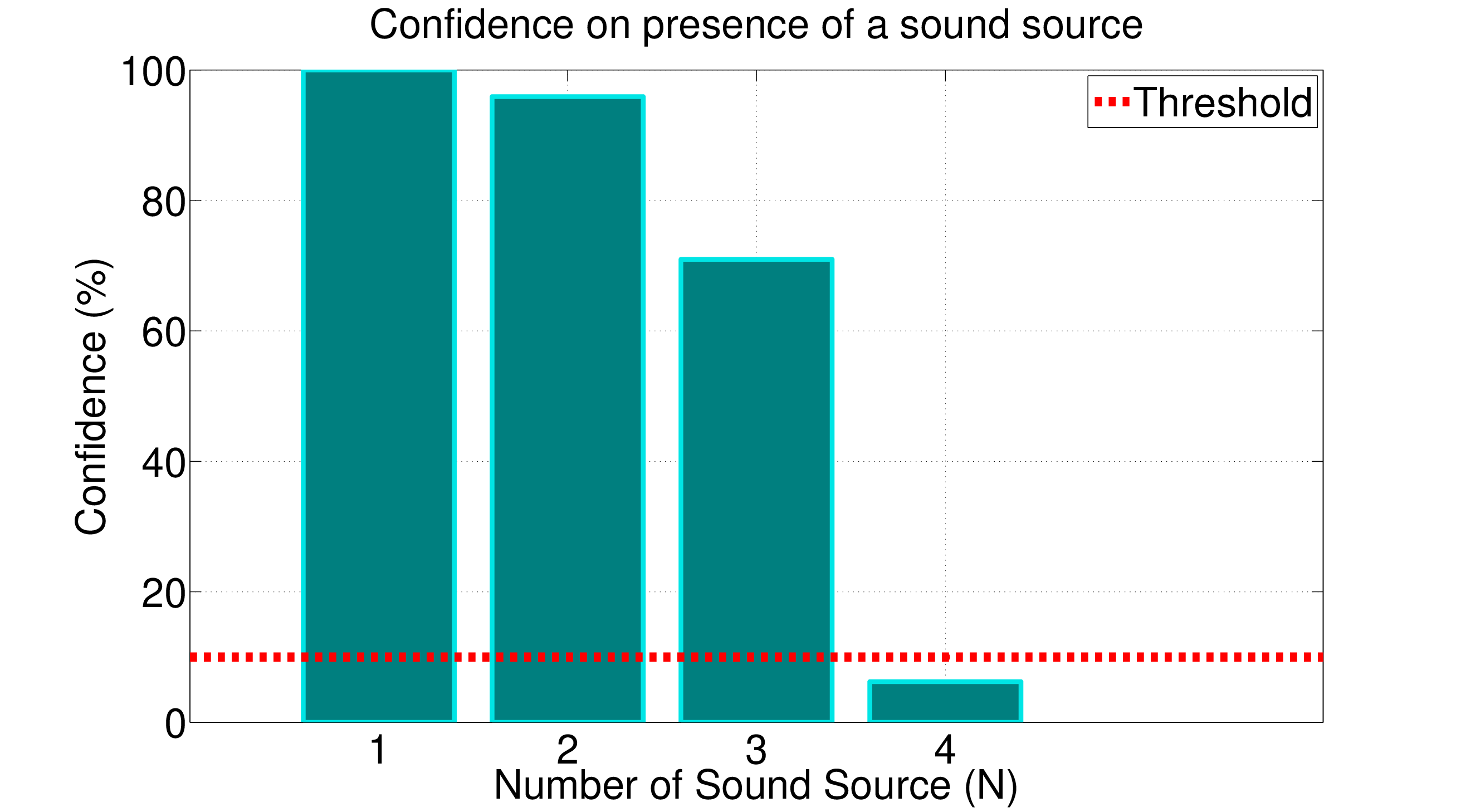}\hfill{}

\caption{Confidence on presence of sound sources. \label{fig:Confidence-Sound-Source}}
\end{figure}

The number of sound sources is determined by carefully selecting a
threshold, as shown in Fig.~\ref{fig:Confidence-Sound-Source}. The
confidence about the presence of a sound source is dependent on the
$count$ value. The source with the maximum $count$ is considered
to be qualified with 100 \% confidence and the confidence values for
other sources are calculated relatively. The source with a confidence
value less than the threshold is considered to be noise and is not
qualified as a sound source. 

\section{Simulation and Experimental Results \label{sec:Results-and-Discussion}}

\subsection{Simulation and Experimental Setup}

\begin{figure}[h]
\hfill{}\includegraphics[width=0.98\columnwidth]{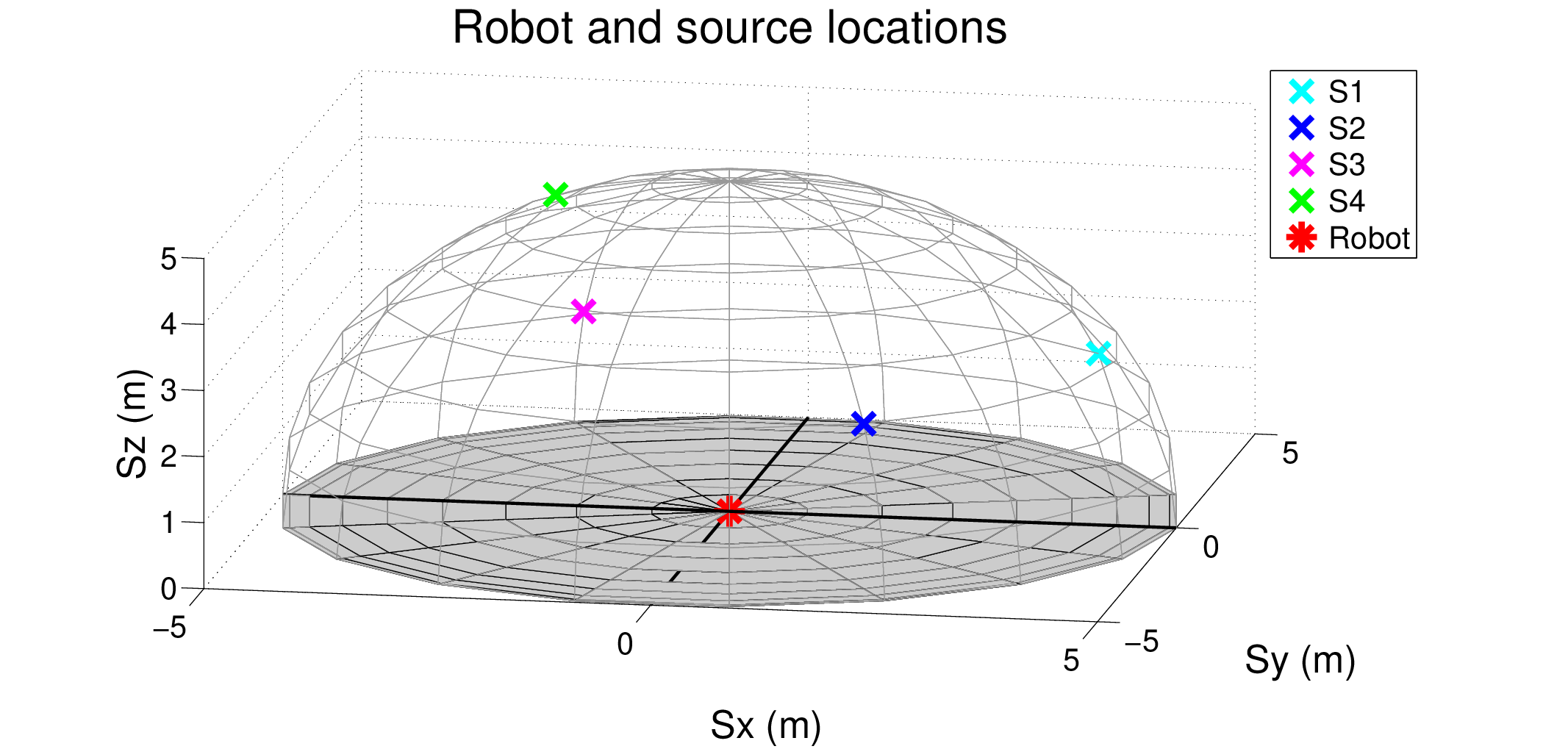}\hfill{}\caption{Simulation setup showing four sound sources placed at $S_{1}(20^{o},\text{ }50^{o})$,
$S_{2}(30^{o},\text{ }150^{o})$, $S_{3}(50^{o},\text{ }200^{o})$
and $S_{4}(60^{o},\text{ }300^{o})$ at a distance of $5$~m from
the center (origin) which is also the center of the microphone array
on the robot. \label{fig:Sound-source-locations}}
\end{figure}

Audio Array Toolbox~\cite{Donohue2009} is used to establish an emulated
rectangular room using the image method described in~\cite{allen1979image}.
The robot was placed in the origin of the room. The sound sources
and the microphones are assumed omnidirectional and the attenuation
of the sound are calculated per the specifications in Table\emph{\large{}~}\ref{tab:Room specifications}.
Fig.~\ref{fig:Sound-source-locations} shows the simulation setup
with the robot placed at the origin and the four sound sources placed
at different azimuth and elevation angles. 
\begin{table}[H]
\caption{Simulated room specifications \label{tab:Room specifications}}

\hfill{}%
\begin{tabular}{|c|c|}
\hline 
Parameter & Value\tabularnewline
\hline 
\hline 
Dimension & 20 m x 20 m x 20 m\tabularnewline
\hline 
Reflection coefficient & \multirow{2}{*}{0.5}\tabularnewline
(walls, floor and ceiling) & \tabularnewline
\hline 
Sound speed & 345 m/s\tabularnewline
\hline 
Temperature & 22$^{o}C$\tabularnewline
\hline 
Static pressure & 29.92 mmHg\tabularnewline
\hline 
Relative humidity & 38 \%\tabularnewline
\hline 
\end{tabular}\hfill{}
\end{table}
 
A number of recorded audio signals available at~\cite{Corpus} were
used as sound sources to test the technique. Different numbers of
sound sources were placed at various azimuth and elevation angles
at a fixed distance of $5\,\text{m}$ and the ICTD signal was recorded
by the rotating bi-microphone array with mics separated by a distance
of $0.18$ m. The sound sources were separated by at least $20^{o}$
in azimuth and at least $10^{o}$ in elevation. The ICTD value was
calculated and recorded every $1^{o}$ of rotation. Zero-mean noise
with a variance ($\sigma_{noise})$ of $0.001$ was added to this
ICTD signal in simulations to account for sensor noise.
\begin{figure}[h]
\hfill{}\includegraphics[width=0.5\columnwidth]{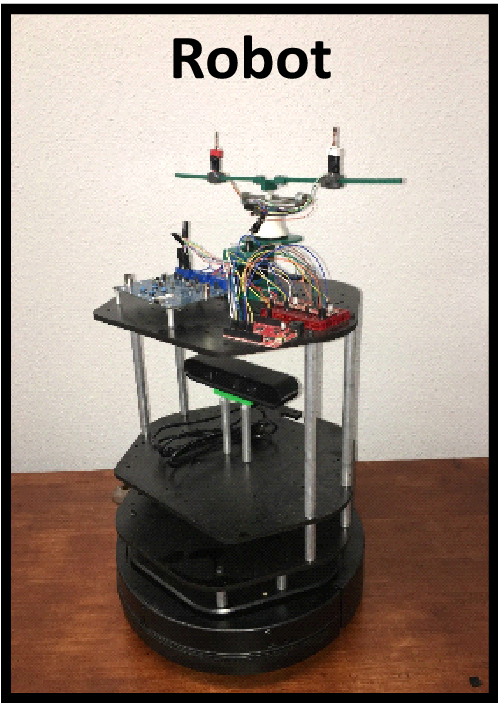}\hfill{}

\caption{Kobuki Turtlebot robot with the platform mounted with two microelectromechanical
systems (MEMS) microphones.\label{fig:Robot}}
\end{figure}
\begin{figure}[h]
\hfill{}\includegraphics[width=0.98\columnwidth]{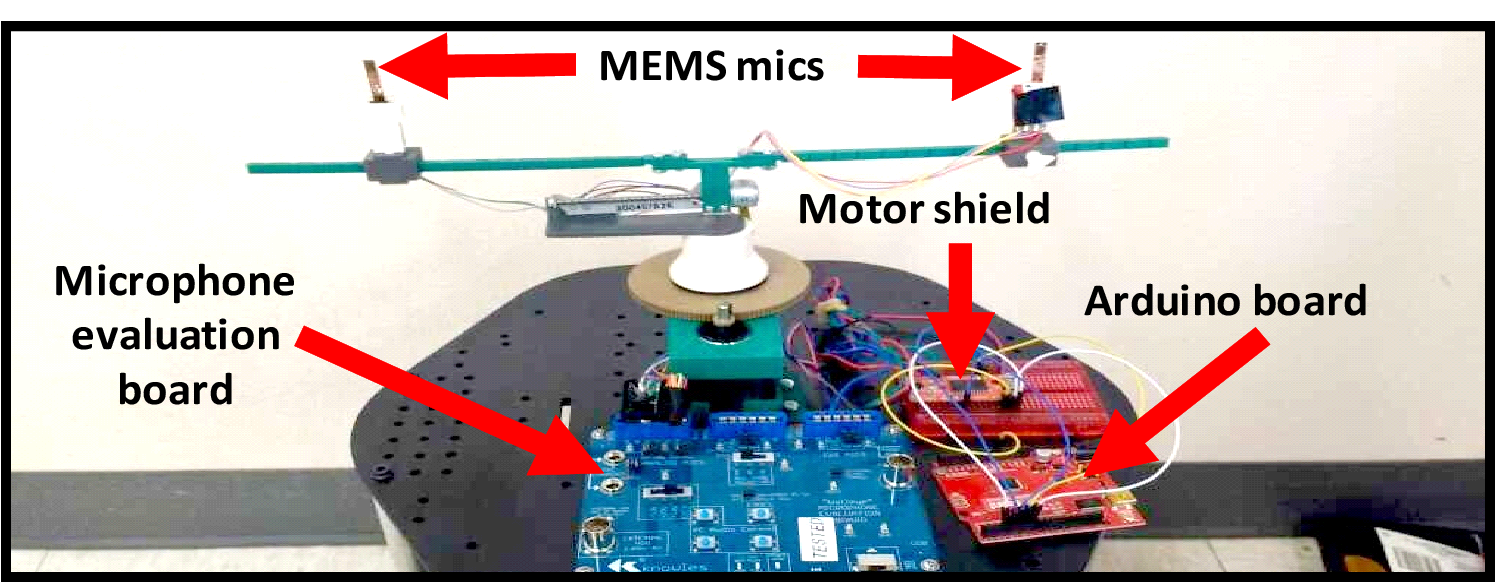}\hfill{}

\caption{Robotic platform with two microelectromechanical systems (MEMS) microphones,
microphone evaluation board for acquisition of sound signals recorded
by the microphones, motor shield and the arduino board used to rotate
the bipolar stepper motor.\label{fig:Robotic_platform}}
\end{figure}
\begin{figure}[h]
\hfill{}\includegraphics[width=0.5\columnwidth]{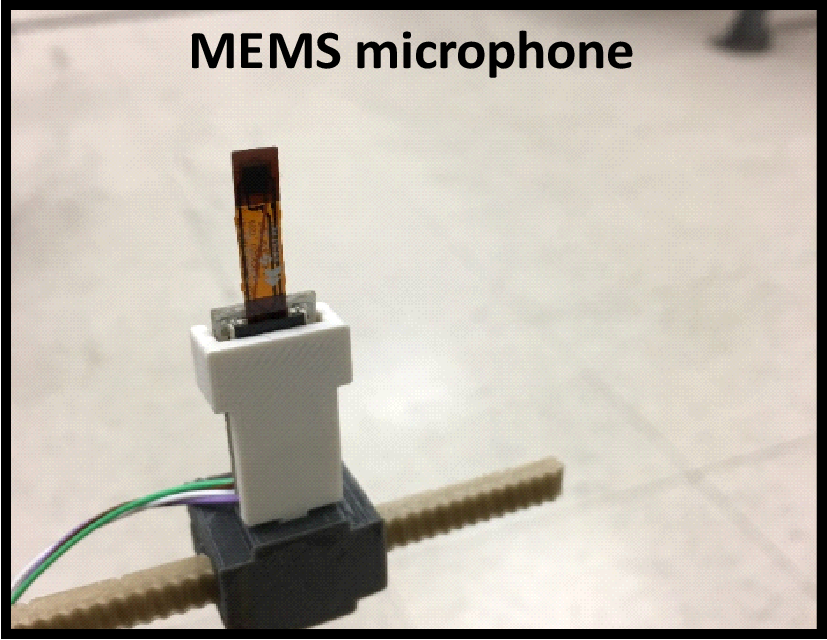}\hfill{}

\caption{One of the two microelectromechanical systems (MEMS) microphone mounted
on the robotic platform.\label{fig:MEMS_mic}}
\end{figure}
\begin{figure}[h]
\hfill{}\includegraphics[width=0.98\columnwidth]{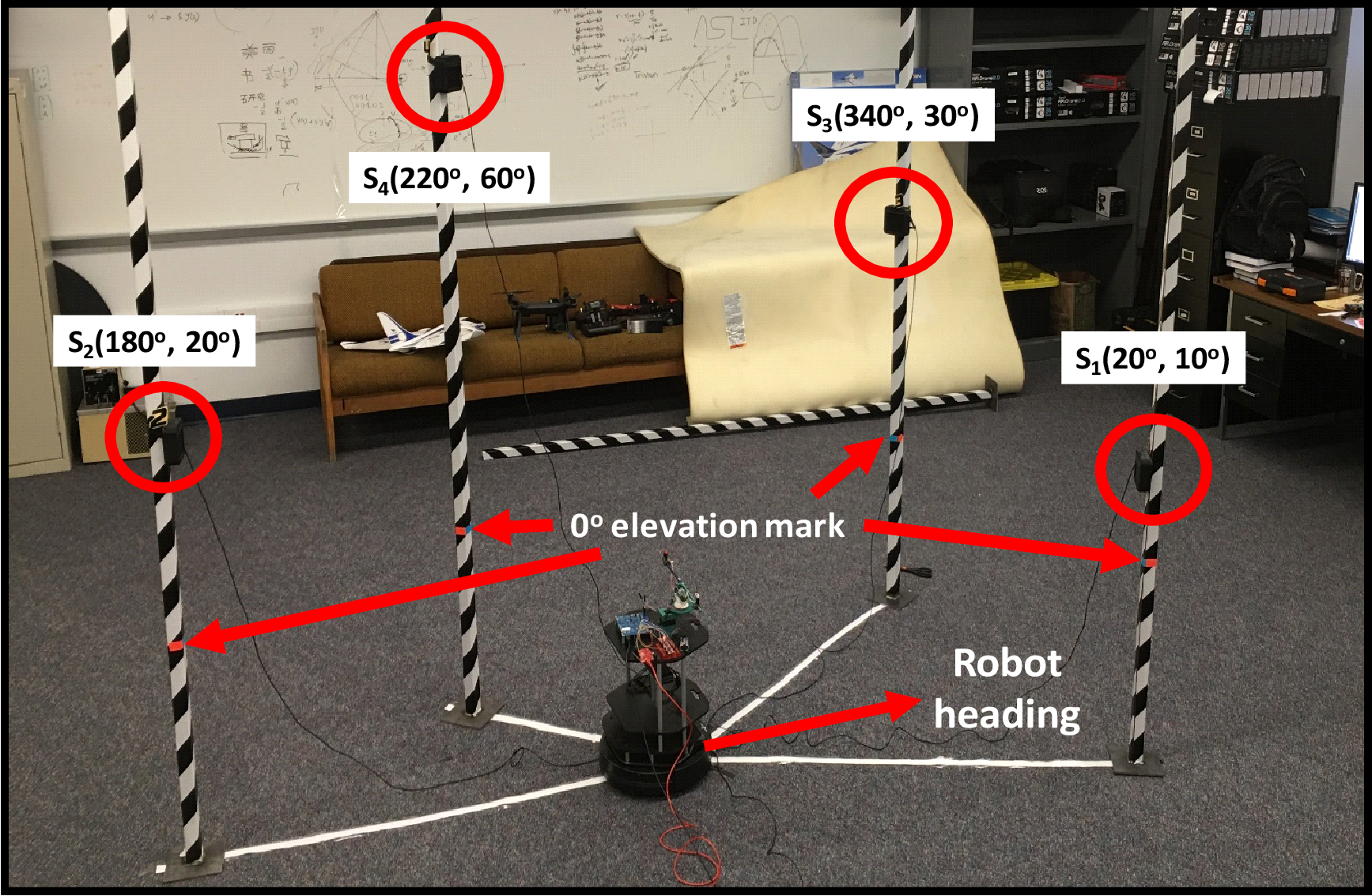}\hfill{}

\caption{Experimental setup showing four sound sources mounted on the four
poles. The poles are marked with the zero-degree elevation which corresponds
to the height of the bi-microphone array on the robotic platform placed
at the center (origin).\label{fig:Experimental-setup}}
\end{figure}

\begin{figure}[h]
\hfill{}\includegraphics[width=0.98\columnwidth]{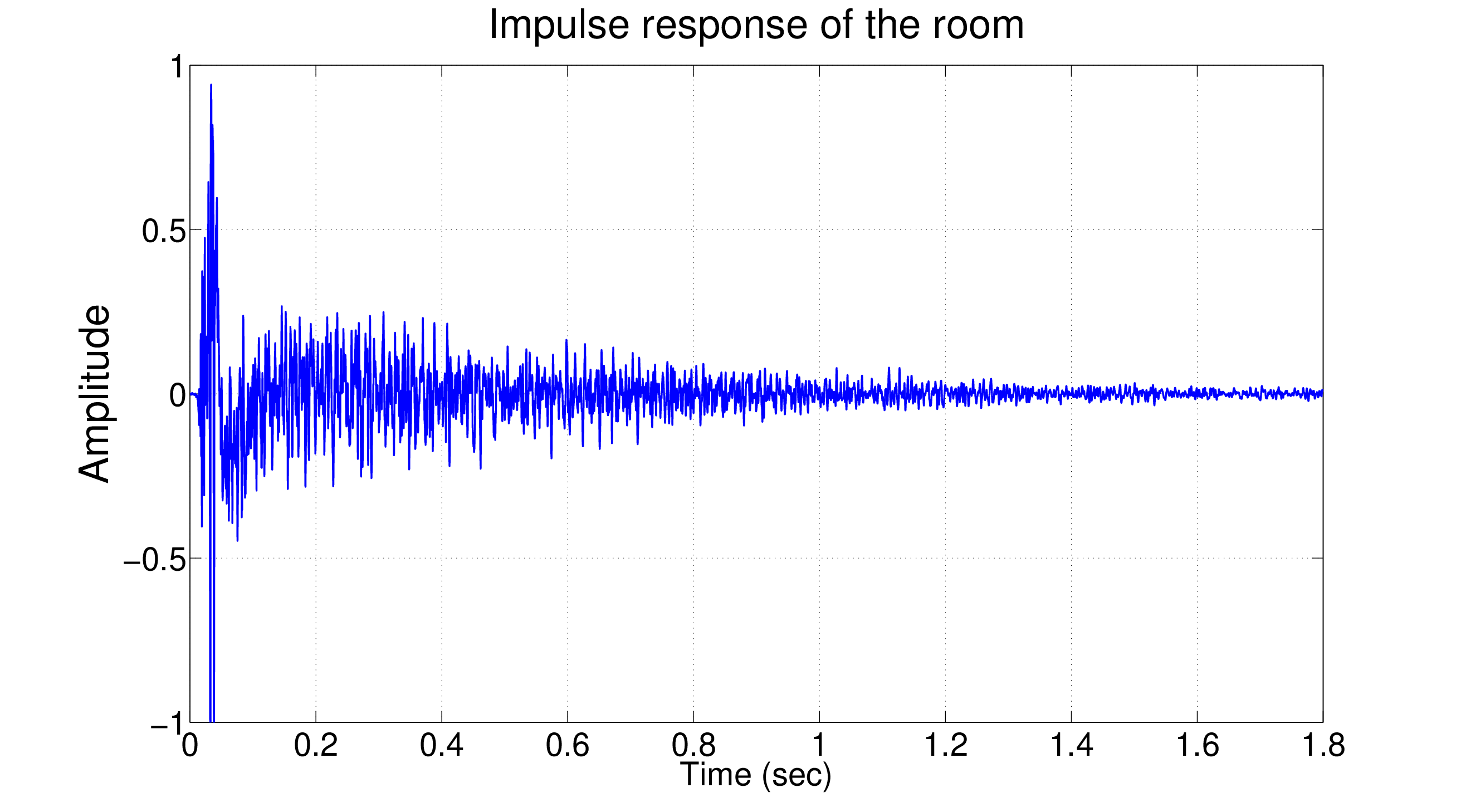}\hfill{}

\caption{Impulse response of the room reverberation showing secondary peaks
representing the reflections from the floor and the walls. \label{fig:Impulse-response}}
\end{figure}

Experiments were conducted using a Kobuki Turtlebot 2 robot shown
in Fig.~\ref{fig:Robot} mounted with a robotic platform as shown
in Fig.~\ref{fig:Robotic_platform} consisting of two microelectromechanical
systems (MEMS) analog microphones shown in Fig.~\ref{fig:MEMS_mic}.
These experiments were conducted in an indoor environment with the
reverberation time $RT60=670$~ms (where RT60 is the time required
for a sound to decay 60 dB). Fig.~\ref{fig:Impulse-response} shows
the impulse response of the room. The experimental setup is shown
in Fig.~\ref{fig:Experimental-setup}. The sampling frequency of
the two microphones was 44,100 Hz while recording the signal. A microphone
evaluation board assembly was used for data acquisition. The angular
speed of the rotation of the microphone array was controlled by a
bipolar stepper motor with gear ratio adjusted to $0.9^{o}$ per step
and rotating at an angular velocity of $2\pi/5$ rad/sec which was
controlled by an Arduino board. The distance between the two microphones
was kept constant as $0.3$ m. Audio clips were played in a number
of loudspeakers which were used as sound sources. These speakers were
kept at different locations with a $1.5$ m distance from the robot,
thereby satisfying the far-field approximation~\cite{Calmes2009}.
\begin{table}[H]
\caption{Parameters for RANSAC-Based and DBSCAN-Based MSSL \label{tab:Parameters-for-RANSAC-and-DBSCAN}}

\hfill{}%
\begin{tabular}{|c|c|c|}
\hline 
Parameters & For simulations & For experiments\tabularnewline
\hline 
\hline 
$\sigma_{conf}$ & $0.0157$ m & $0.0261$ m\tabularnewline
\hline 
$N_{D}$ & $10000$ & $10000$\tabularnewline
\hline 
$N_{R}$ & $5000$ & $5000$\tabularnewline
\hline 
Threshold & $10$ \% & $7$ \%\tabularnewline
\hline 
$\epsilon$ & $3^{o}$ & $3^{o}$\tabularnewline
\hline 
$m$ & $40$ & $40$\tabularnewline
\hline 
$\sigma_{noise}$ & $0.001$ m & \textendash{}\tabularnewline
\hline 
\end{tabular}\hfill{}
\end{table}

\subsection{Results and Discussion}

\begin{figure}[H]
\subfloat[Simulation results showing mapped data points for three sound sources
placed at $(30^{o},340^{o})$, $(10^{o},20^{o})$, and $(20^{o},180^{o})$.
\label{fig:Mapped-points}]{\hfill{}\includegraphics[width=0.98\columnwidth]{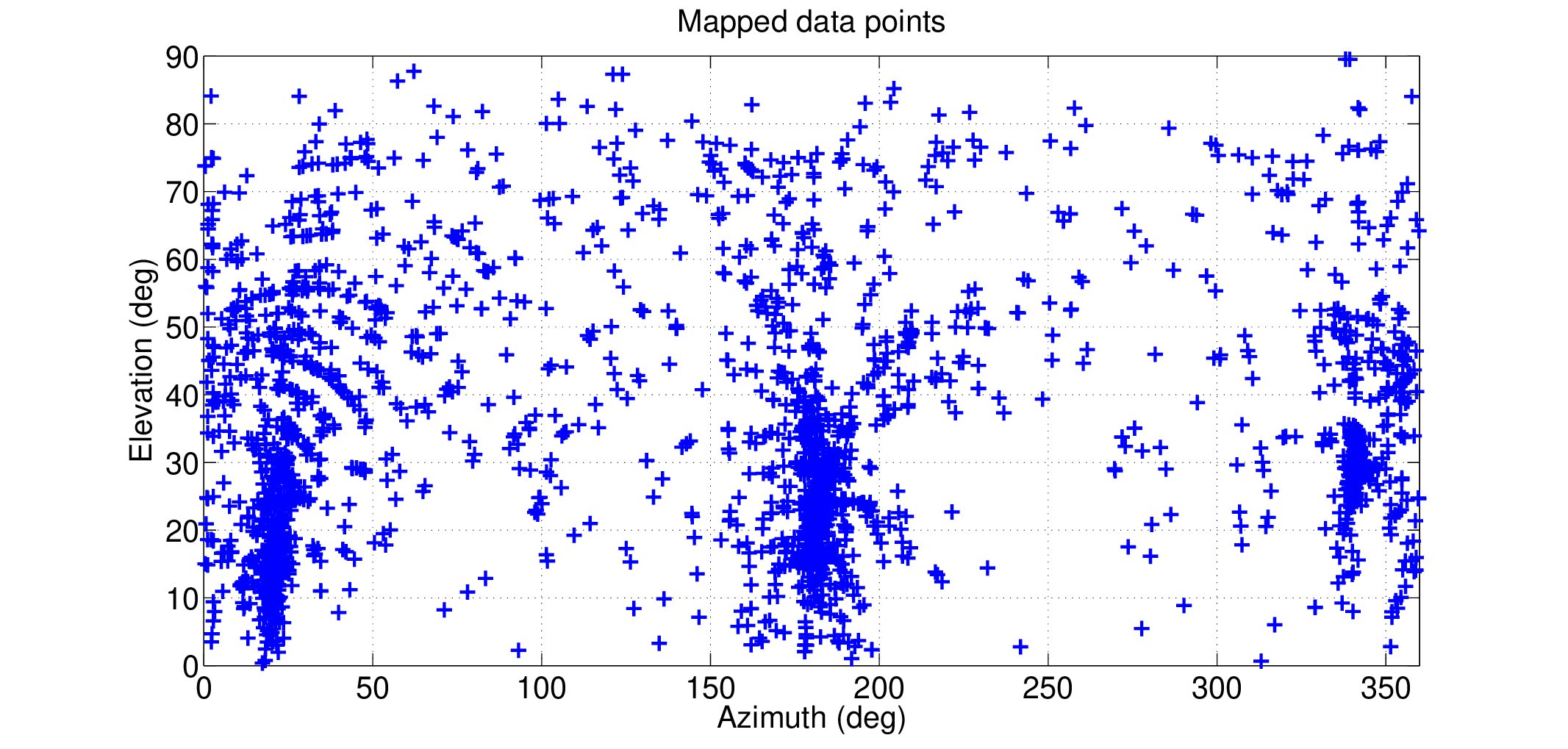}\hfill{}

}

\subfloat[Experimental results showing mapped data points for four sound sources
placed at $(30^{o},340^{o})$, $(10^{o},20^{o})$, $(60^{o},220^{o})$
and $(20^{o},180^{o})$. \label{fig:Mapped-points-Hardware}]{\hfill{}\includegraphics[width=0.98\columnwidth]{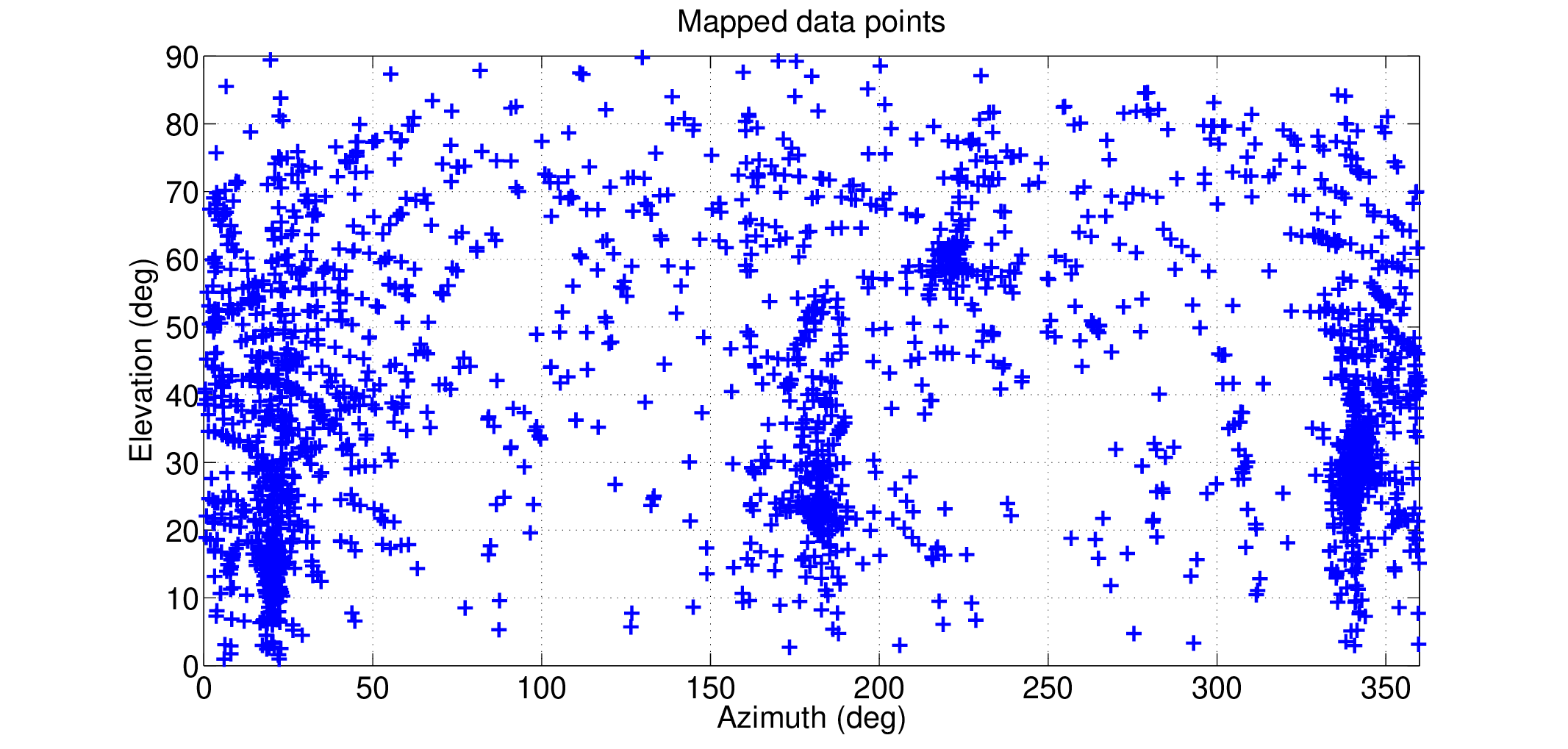}\hfill{}}\caption{ICTD samples mapped to the orientation domain.}
\end{figure}
Figures ~\ref{fig:Mapped-points} and \ref{fig:Mapped-points-Hardware}
show the formation of the clusters in simulation and real environment,
respectively, during the mapping of ICTD data points to their respective
$\varphi$ and $\theta$ angles using Equations ~\ref{eq:Theta-from-A},
~\ref{eq:Amplitude-two-points} and ~\ref{eq:Azimuth-two-points}.
\begin{figure}[H]
\subfloat[Simulation results showing clusters detected using DBSCAN-based MSSL
technique. Three sound sources placed at $(30^{o},340^{o})$, $(10^{o},20^{o})$,
and $(20^{o},180^{o})$ which were localized at $(29.91^{o},340.11^{o})$,
$(14.40^{o},20.01^{o})$ and $(21.02^{o},180.58^{o})$ respectively.
\label{fig:DBSCAN_clusters_Simulation}]{\hfill{}\includegraphics[width=0.98\columnwidth]{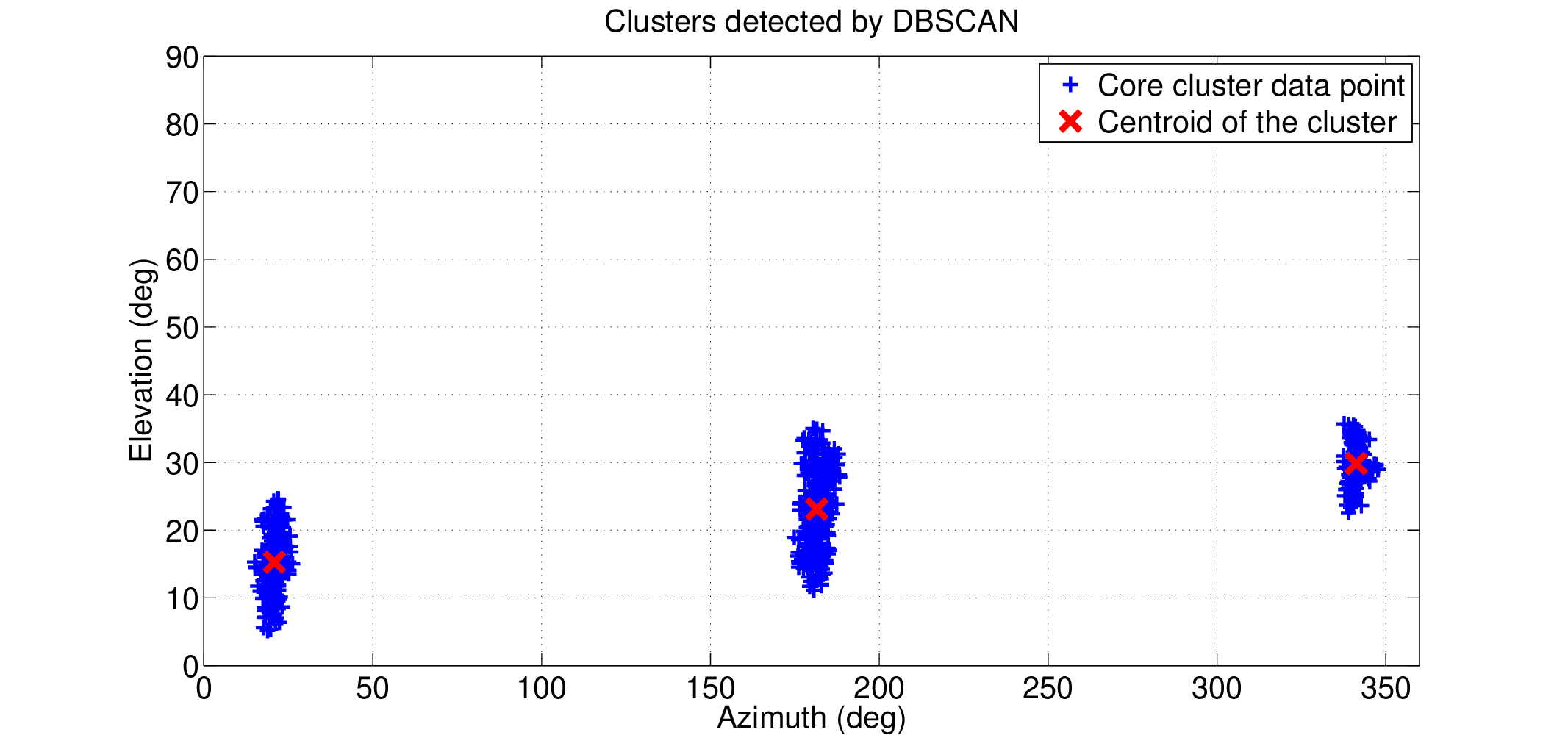}\hfill{}}

\subfloat[Experimental results showing clusters detected using DBSCAN-based
MSSL technique. Four sound sources placed at $(30^{o},340^{o})$,
$(10^{o},20^{o})$, $(60^{o},220^{o})$ and $(20^{o},180^{o})$ in
the real environment which were localized at $(29.35^{o},341.06^{o})$,
$(13.55^{o},20.36^{o})$, $(60.20^{o},220.76^{o})$ and $(22.87^{o},182.05^{o})$
respectively. \label{fig:DBSCAN_clusters-Hardware}]{\hfill{}\includegraphics[width=0.98\columnwidth]{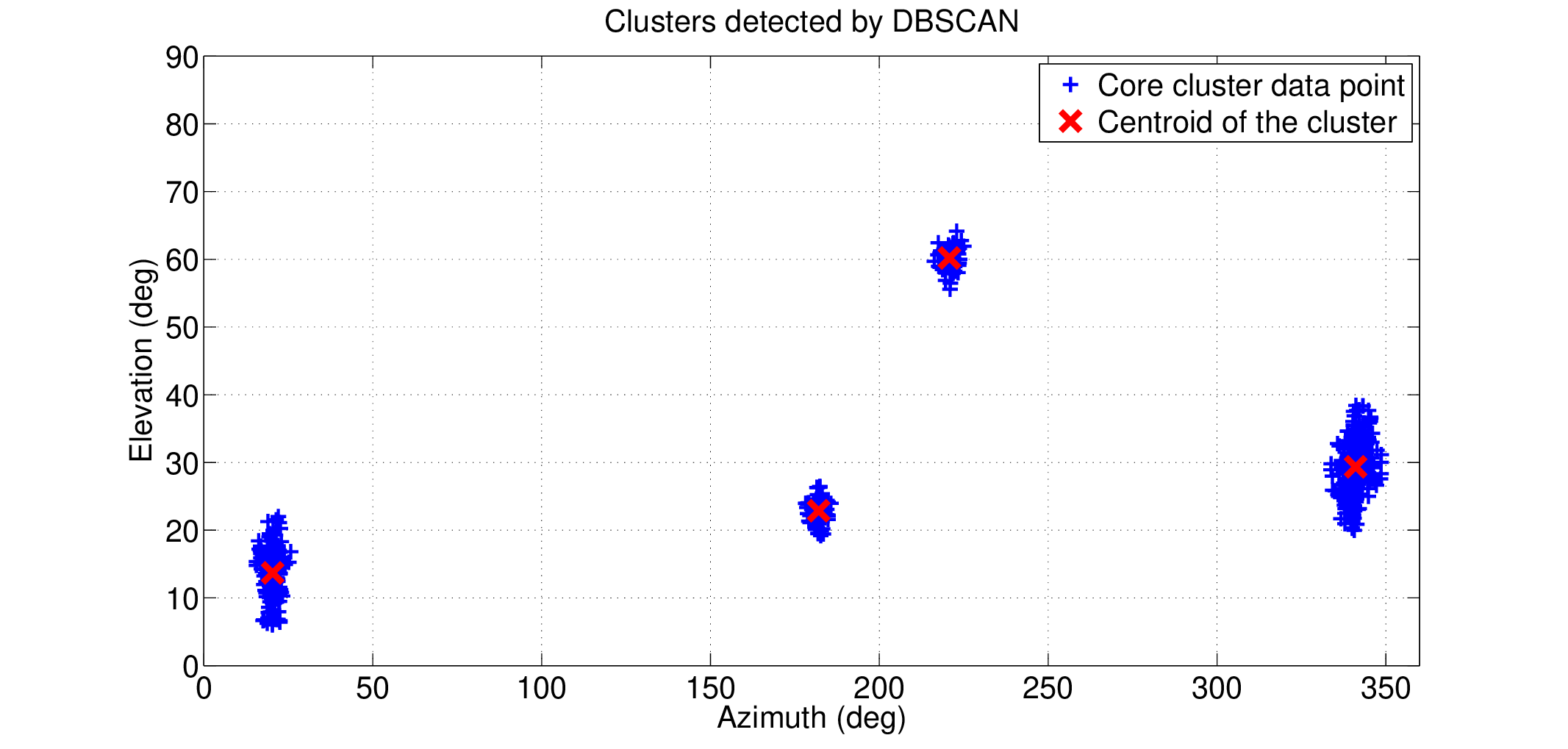}\hfill{}}\caption{Clusters detected using DBSCAN technique}
\end{figure}
 Figures ~\ref{fig:DBSCAN_clusters_Simulation} and \ref{fig:DBSCAN_clusters-Hardware}
show the detection of these clusters in simulation and real environment,
respectively, using the DBSCAN technique for a sample run.
\begin{table}[h]
\caption{Mean absolute error (MAE) for localization performed with DBSCAN-Based
and RANSAC-Based MSSL in simulation (Sim) and experiments (Expt) for
different number of sound sources. \label{tab:Results-DBSCAN}}

\hfill{}%
\begin{tabular}{|c|c|c|c|c|}
\hline 
\multicolumn{1}{|c|}{Number} & \multicolumn{2}{c|}{MAE (Sim)} & \multicolumn{2}{c|}{MAE (Expt)}\tabularnewline
\cline{2-5} \cline{3-5} \cline{4-5} \cline{5-5} 
of & $\varphi$ (deg) & $\theta$ (deg) & $\varphi$ (deg) & $\theta$ (deg)\tabularnewline
\cline{2-5} \cline{3-5} \cline{4-5} \cline{5-5} 
source(s) & \multicolumn{4}{c|}{DBSCAN}\tabularnewline
\hline 
4 & 1.73 & 3.20 & 3.71 & 5.66\tabularnewline
\hline 
3 & 0.8 & 4.68 & 1.77 & 5.78\tabularnewline
\hline 
2 & 1.06 & 2.18 & 2.89 & 3.92\tabularnewline
\hline 
1 & 0.94 & 0.57 & 2.27 & 0.35\tabularnewline
\hline 
 & \multicolumn{4}{c|}{RANSAC}\tabularnewline
\hline 
4 & 0.88 & 8.12 & 3.92 & 7.20\tabularnewline
\hline 
3 & 2.51 & 5.56 & 2.33 & 5.62\tabularnewline
\hline 
2 & 2.55 & 5.01 & 3.07 & 3.88\tabularnewline
\hline 
1 & 1.61 & 1.97 & 2.15 & 3.07\tabularnewline
\hline 
\end{tabular}\hfill{}
\end{table}
The parameters used for the RANSAC- and DBSCAN-based algorithms are
listed in Table~\ref{tab:Parameters-for-RANSAC-and-DBSCAN}. The
value $\sigma_{conf}$ was chosen to be $0.0157$~m for simulation
and $0.0261$~m for experiments, which implies that the sound sources
with the same azimuth are assumed to be separated by at least $5^{o}$
in elevation. Table~\ref{tab:Results-DBSCAN} shows the simulation
and experimental results of localization with the number of sound
sources varying from one to four.

$1000$ Monte Carlo simulation runs were performed using the two proposed
approaches, respectively, with specifications given in Table~\ref{tab:Room specifications}.
Simulations were run with $K=1,2,3,4,5$ sources and the results of
the source counting are listed in the Table~\ref{tab:Source-count-using-DBSCAN}.
\begin{table}[H]
\caption{Estimated vs actual number of sound source count for DBSCAN-Based
and RANSAC-Based MSSL in the simulated environment. \label{tab:Source-count-using-DBSCAN}}
\hfill{}%
\begin{tabular}{|c|c|c|c|c|c|c|c|c|}
\hline 
\multirow{13}{*}{Act} & K & 1 & 2 & 3 & 4 & 5 & 6 & $\geq$ 7\tabularnewline
\cline{2-9} \cline{3-9} \cline{4-9} \cline{5-9} \cline{6-9} \cline{7-9} \cline{8-9} \cline{9-9} 
 &  & \multicolumn{7}{c|}{Estimated - DBSCAN}\tabularnewline
\cline{2-9} \cline{3-9} \cline{4-9} \cline{5-9} \cline{6-9} \cline{7-9} \cline{8-9} \cline{9-9} 
 & 1 & 944 & 56 & 0 & 0 & 0 & 0 & 0\tabularnewline
\cline{2-9} \cline{3-9} \cline{4-9} \cline{5-9} \cline{6-9} \cline{7-9} \cline{8-9} \cline{9-9} 
 & 2 & 11 & 902 & 65 & 15 & 7 & 0 & 0\tabularnewline
\cline{2-9} \cline{3-9} \cline{4-9} \cline{5-9} \cline{6-9} \cline{7-9} \cline{8-9} \cline{9-9} 
 & 3 & 1 & 61 & 847 & 47 & 38 & 5 & 1\tabularnewline
\cline{2-9} \cline{3-9} \cline{4-9} \cline{5-9} \cline{6-9} \cline{7-9} \cline{8-9} \cline{9-9} 
 & 4 & 12 & 39 & 126 & 687 & 82 & 36 & 18\tabularnewline
\cline{2-9} \cline{3-9} \cline{4-9} \cline{5-9} \cline{6-9} \cline{7-9} \cline{8-9} \cline{9-9} 
 & 5 & 4 & 17 & 80 & 183 & 595 & 110 & 11\tabularnewline
\cline{2-9} \cline{3-9} \cline{4-9} \cline{5-9} \cline{6-9} \cline{7-9} \cline{8-9} \cline{9-9} 
 &  & \multicolumn{7}{c|}{Estimated - RANSAC}\tabularnewline
\cline{2-9} \cline{3-9} \cline{4-9} \cline{5-9} \cline{6-9} \cline{7-9} \cline{8-9} \cline{9-9} 
 & 1 & 1000 & 0 & 0 & 0 & 0 & 0 & 0\tabularnewline
\cline{2-9} \cline{3-9} \cline{4-9} \cline{5-9} \cline{6-9} \cline{7-9} \cline{8-9} \cline{9-9} 
 & 2 & 7 & 991 & 2 & 0 & 0 & 0 & 0\tabularnewline
\cline{2-9} \cline{3-9} \cline{4-9} \cline{5-9} \cline{6-9} \cline{7-9} \cline{8-9} \cline{9-9} 
 & 3 & 0 & 56 & 898 & 46 & 0 & 0 & 0\tabularnewline
\cline{2-9} \cline{3-9} \cline{4-9} \cline{5-9} \cline{6-9} \cline{7-9} \cline{8-9} \cline{9-9} 
 & 4 & 0 & 4 & 104 & 888 & 4 & 0 & 0\tabularnewline
\cline{2-9} \cline{3-9} \cline{4-9} \cline{5-9} \cline{6-9} \cline{7-9} \cline{8-9} \cline{9-9} 
 & 5 & 0 & 0 & 1 & 85 & 750 & 139 & 25\tabularnewline
\hline 
\end{tabular}\hfill{}
\end{table}

\begin{figure}
\subfloat[Simulation results of estimation of signal $d_{n}$ from the multi-source
signal $d$ using RANSAC-based MSSL technique. Three sound sources
placed at $(30^{o},340^{o})$, $(10^{o},20^{o})$, and $(20^{o},180^{o})$
which were localized at $(33.11^{o},338.86^{o})$, $(14.40^{o},22.01^{o})$
and $(24.27^{o},181.15^{o})$ respectively. \label{fig:Mult-Source-ITD-Estimation}]{\hfill{}\includegraphics[width=0.98\columnwidth]{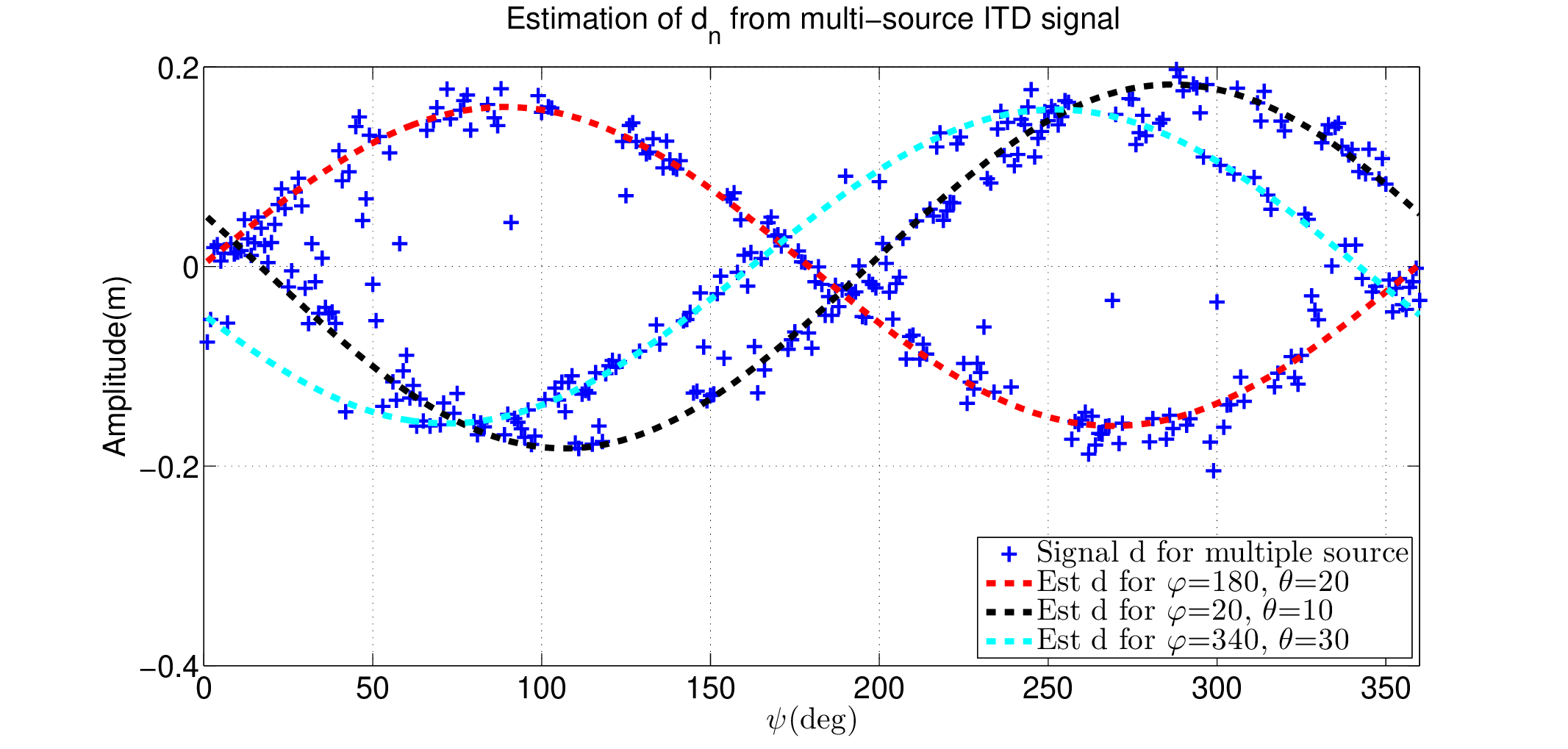}\hfill{}}

\subfloat[Experimental results of estimation of signal $d_{n}$ from the multi-source
signal $d$ using RANSAC-based MSSL technique. Four sound sources
placed at $(30^{o},340^{o})$, $(10^{o},20^{o})$, $(60^{o},220^{o})$
and $(20^{o},180^{o})$ which were localized at $(32.41^{o},341.26^{o})$,
$(13.69^{o},20.86^{o})$, $(58.20^{o},219.02^{o})$ and $(22.27^{o},181.15^{o})$
respectively. \label{fig:Mult-Source-ITD-Estimation-RoboticPlatform}]{\hfill{}\includegraphics[width=0.98\columnwidth]{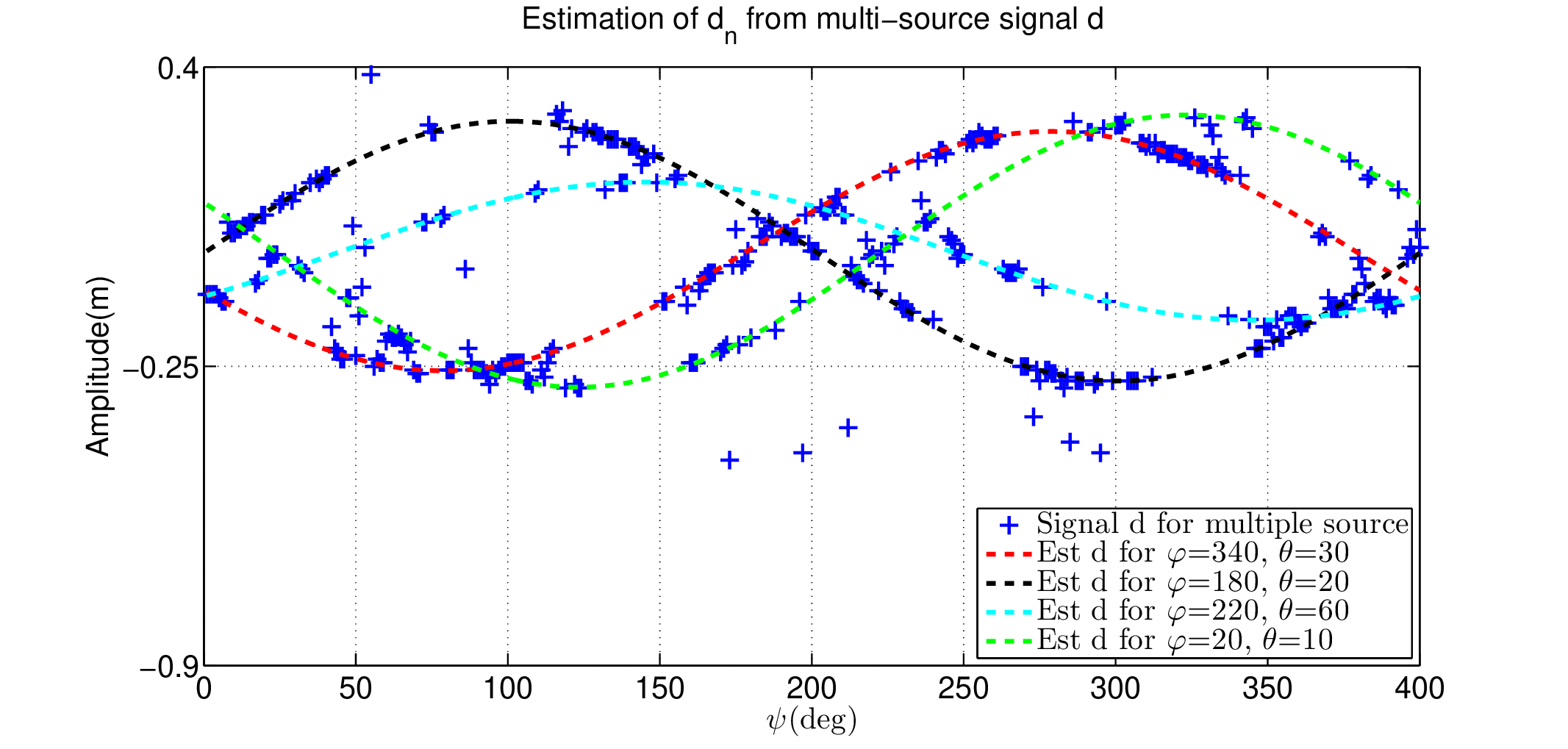}\hfill{}}\caption{Estimated signals $d_{n}$ from the multi-source ICTD signal $d$ }
\end{figure}

\begin{figure}[h]
\hfill{}\includegraphics[width=0.98\columnwidth]{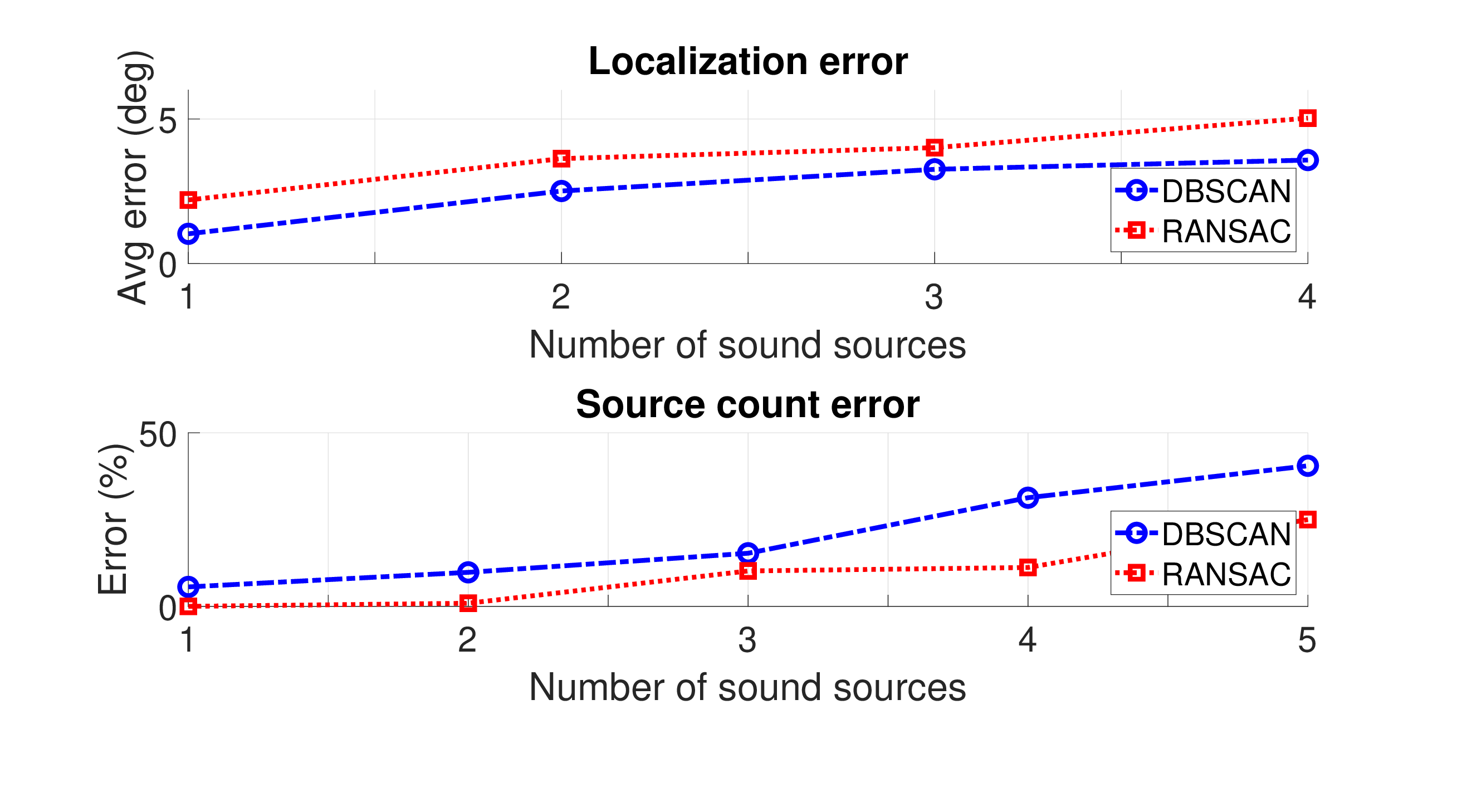}\hfill{}

\caption{Top subfigure represents the average of simulation and experimental
localization error and the bottom subfigure represents the number
of sound source identification percentage error in simulation by the
DBSCAN-based and RANSAC-based MSSL for different number of sound sources.
\label{fig:LocalizationSourceCount-error}}
\end{figure}

Figure ~\ref{fig:Mult-Source-ITD-Estimation} and \ref{fig:Mult-Source-ITD-Estimation-RoboticPlatform}
shows the result of a sample run by the RANSAC-based algorithm in
simulation and real environment respectively. 
 Since the ICTD signal is noisy, any point very close ($\sigma_{conf}=0.015688$)
to any of $d_{n}$ was chosen to be on the ICTD by the RANSAC algorithm.
The signal to noise ratio (SNR) of the measured signal $d$ was $18.94$
dB. For a source to be considered as a qualified sound source, the
threshold for the confidence that worked for us was $10$\% in simulation
and $7$\% in experiments. 
\begin{figure}
\subfloat[Experimental results for confidence on presence of sound sources placed
at $(30^{o},340^{o})$, $(10^{o},20^{o})$, $(60^{o},220^{o})$ and
$(20^{o},180^{o})$ using DBSCAN. \label{fig:Confidence-Sound-Source-DBSCAN}]{\hfill{}\includegraphics[width=0.98\columnwidth]{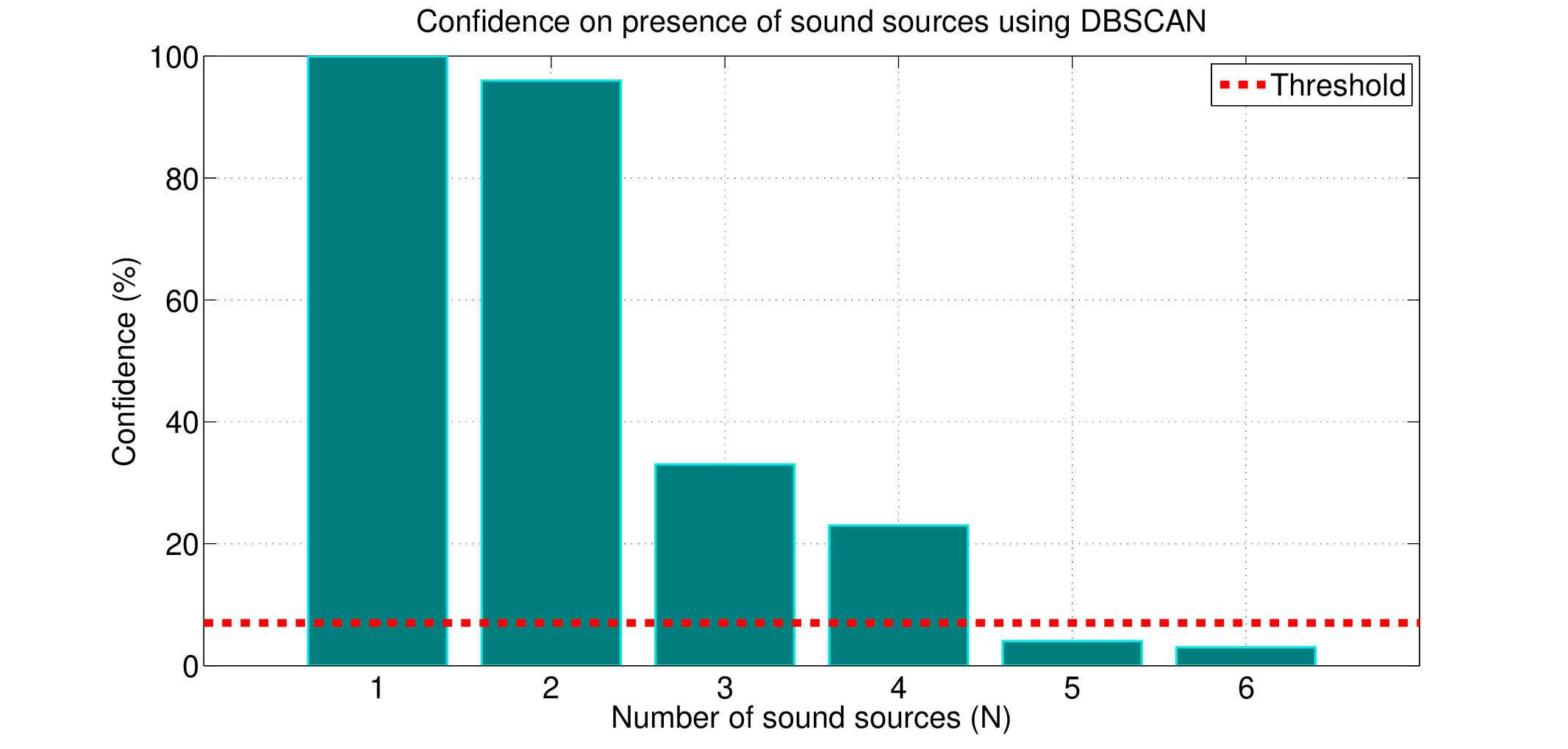}\hfill{}

}

\subfloat[Experimental results for confidence on presence of sound sources placed
at $(30^{o},340^{o})$, $(10^{o},20^{o})$, $(60^{o},220^{o})$ and
$(20^{o},180^{o})$ using RANSAC. \label{fig:Confidence-Sound-Source-RANSAC}]{\hfill{}\includegraphics[width=0.98\columnwidth]{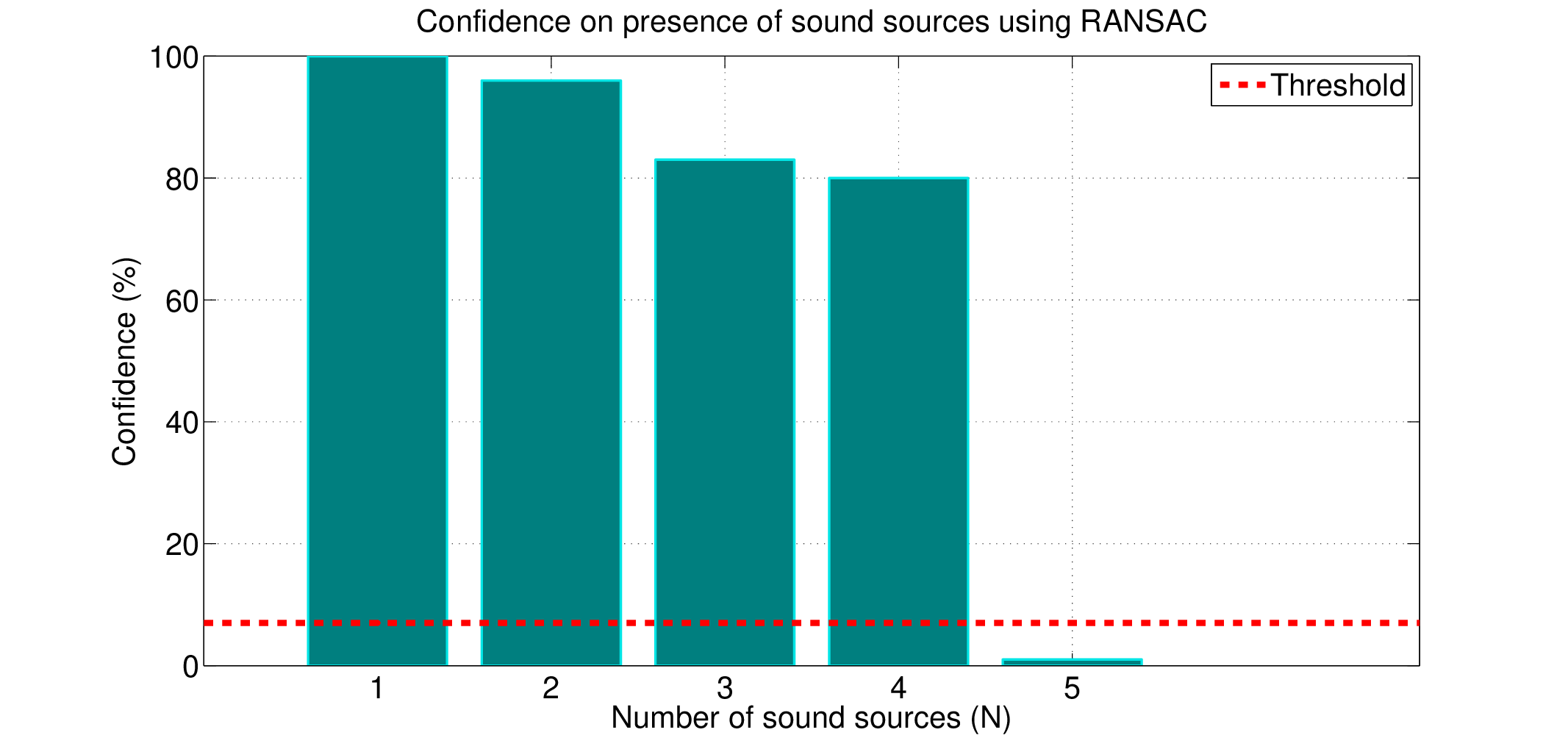}\hfill{}

}\caption{Experimental results for source count}
\end{figure}
As shown in top subfigure of Fig.~\ref{fig:LocalizationSourceCount-error},
the average error of orientation localization with the DBSCAN-based
algorithm is less as compared to the RANSAC-based algorithm, which,
however, generates comparatively more accurate results for source
counting, as shown in the bottom subfigure of Fig.~\ref{fig:LocalizationSourceCount-error}.
In both simulations and experiments, the error of elevation angle
estimation was found to be large for sources kept close to zero elevation,
which coincides the conclusion in~\cite{Gala2018}. The performance
of the localization and source counting using both proposed techniques
can be improved by increasing the number of rotations of the bi-microphone
array. 

\section{Conclusion and Future Scope\label{sec:Conclusion-and-Future-Scope}}

Two novel techniques are presented for small autonomous unmanned vehicles
(SAUVs) to perform 3D multi-sound-source localization (MSSL) using
a self-rotating bi-microphone array. The Density-Based Spatial Clustering
of Applications with Noise (DBSCAN) based MSSL approach iteratively
maps the randomly chosen points in the inter-channel time difference
(ICTD) signal to the orientation domain, leading to a data sets for
clustering. The number of clusters represents the number of sound
sources and the location of the centroid of a cluster represents the
location of a sound source. The Random Sample Consensus (RANSAC) based
approach iteratively estimates parameters of a model using two randomly
chosen data points from the ICTD signal data. It then uses a threshold
to decide the number of qualified sound sources. The simulation and
experimental results show the effectiveness of both approaches in
identifying the number and the 3D orientations of the sound sources.

The techniques presented in the paper are able to localize multiple
stationary sound sources by a stationary robotic platform. Considerations
on the presence of obstacles and the robot/sound motions motivate
our future work.

\bibliographystyle{IEEEtran}
\bibliography{References}

\begin{thebibliography}{10}
\providecommand{\url}[1]{#1}
\csname url@rmstyle\endcsname
\providecommand{\newblock}{\relax}
\providecommand{\bibinfo}[2]{#2}
\providecommand\BIBentrySTDinterwordspacing{\spaceskip=0pt\relax}
\providecommand\BIBentryALTinterwordstretchfactor{4}
\providecommand\BIBentryALTinterwordspacing{\spaceskip=\fontdimen2\font plus
\BIBentryALTinterwordstretchfactor\fontdimen3\font minus
  \fontdimen4\font\relax}
\providecommand\BIBforeignlanguage[2]{{%
\expandafter\ifx\csname l@#1\endcsname\relax
\typeout{** WARNING: IEEEtran.bst: No hyphenation pattern has been}%
\typeout{** loaded for the language `#1'. Using the pattern for}%
\typeout{** the default language instead.}%
\else
\language=\csname l@#1\endcsname
\fi
#2}}

\bibitem{Wang2016}
Q.~Wang, K.~Ren, M.~Zhou, T.~Lei, D.~Koutsonikolas, and L.~Su, ``Messages
  behind the sound: real-time hidden acoustic signal capture with
  smartphones,'' in \emph{Proceedings of the 22nd Annual International
  Conference on Mobile Computing and Networking}.\hskip 1em plus 0.5em minus
  0.4em\relax ACM, 2016, pp. 29--41.

\bibitem{Boehme2003}
H.-J. B{\"o}hme, T.~Wilhelm, J.~Key, C.~Schauer, C.~Schr{\"o}ter, H.-M.
  Gro{\ss}, and T.~Hempel, ``An approach to multi-modal human--machine
  interaction for intelligent service robots,'' \emph{Robotics and Autonomous
  Systems}, vol.~44, no.~1, pp. 83--96, 2003.

\bibitem{Murray2004}
J.~C. Murray, H.~Erwin, and S.~Wermter, ``Robotics sound-source localization
  and tracking using interaural time difference and cross-correlation,'' in
  \emph{AI Workshop on NeuroBotics}, 2004.

\bibitem{Borenstein1996}
J.~Borenstein, H.~Everett, and L.~Feng, \emph{Navigating mobile robots: systems
  and techniques}.\hskip 1em plus 0.5em minus 0.4em\relax A K Peters Ltd.,
  1996.

\bibitem{Rabinkin1998}
D.~V. {Rabinkin}, ``{Optimum sensor placement for microphone arrays},'' Ph.D.
  dissertation, RUTGERS The State University of New Jersey - New Brunswick,
  1998.

\bibitem{brandstein2013microphone}
M.~Brandstein and D.~Ward, \emph{Microphone arrays: signal processing
  techniques and applications}.\hskip 1em plus 0.5em minus 0.4em\relax Springer
  Science \& Business Media, 2013.

\bibitem{Wallach1939}
H.~Wallach, ``On sound localization,'' \emph{The Journal of the Acoustical
  Society of America}, vol.~10, no.~4, pp. 270--274, 1939.

\bibitem{Lee2015}
S.~Lee, Y.~Park, and Y.-s. Park, ``Three-dimensional sound source localization
  using inter-channel time difference trajectory,'' \emph{International Journal
  of Advanced Robotic Systems}, vol.~12, no.~12, p. 171, 2015.

\bibitem{Handzel2002}
A.~A. Handzel and P.~Krishnaprasad, ``Biomimetic sound-source localization,''
  \emph{IEEE Sensors Journal}, vol.~2, no.~6, pp. 607--616, 2002.

\bibitem{Eriksen2006}
G.~H. Eriksen, ``{Visualization tools and graphical methods for source
  localization and signal separation},'' Master's thesis, {Universityof OSLO,
  Department of Informatics}, 2006.

\bibitem{Zhong2015}
X.~Zhong, W.~Yost, and L.~Sun, ``Dynamic binaural sound source localization
  with {ITD} cues: Human listeners,'' \emph{The Journal of the Acoustical
  Society of America}, vol. 137, no.~4, pp. 2376--2376, 2015.

\bibitem{Gala2018}
D.~Gala, N.~Lindsay, and L.~Sun, ``Three-dimensional sound source localization
  for unmanned ground vehicles with a self-rotational two-microphone array,''
  in \emph{Proceedings of the 5th international conference of control, dynamic
  systems, and robotics (CDSR'18)}, 2018, pp. 104.1 -- 104.11.

\bibitem{Valin2003}
J.-M. Valin, F.~Michaud, J.~Rouat, and D.~L{\'e}tourneau, ``Robust sound source
  localization using a microphone array on a mobile robot,'' in
  \emph{Intelligent Robots and Systems, 2003.(IROS 2003). Proceedings. 2003
  IEEE/RSJ International Conference on}, vol.~2.\hskip 1em plus 0.5em minus
  0.4em\relax IEEE, 2003, pp. 1228--1233.

\bibitem{Sun2014}
L.~Sun and Q.~Cheng, ``Indoor multiple sound source localization using a novel
  data selection scheme,'' in \emph{48th Annual Conference on Information
  Sciences and Systems (CISS)}.\hskip 1em plus 0.5em minus 0.4em\relax IEEE,
  2014, pp. 1--6.

\bibitem{zhong2016active}
X.~Zhong, L.~Sun, and W.~Yost, ``Active binaural localization of multiple sound
  sources,'' \emph{Robotics and Autonomous Systems}, vol.~85, pp. 83--92, 2016.

\bibitem{Blandin2012}
C.~Blandin, A.~Ozerov, and E.~Vincent, ``Multi-source {TDOA} estimation in
  reverberant audio using angular spectra and clustering,'' \emph{Signal
  Processing}, vol.~92, no.~8, pp. 1950--1960, 2012.

\bibitem{Swartling2011}
M.~Swartling, B.~S{\"a}llberg, and N.~Grbi{\'c}, ``Source localization for
  multiple speech sources using low complexity non-parametric source separation
  and clustering,'' \emph{Signal Processing}, vol.~91, no.~8, pp. 1781--1788,
  2011.

\bibitem{Dong2013}
T.~Dong, Y.~Lei, and J.~Yang, ``An algorithm for underdetermined mixing matrix
  estimation,'' \emph{Neurocomputing}, vol. 104, pp. 26--34, 2013.

\bibitem{Yilmaz2004}
O.~Yilmaz and S.~Rickard, ``Blind separation of speech mixtures via
  time-frequency masking,'' \emph{IEEE Transactions on signal processing},
  vol.~52, no.~7, pp. 1830--1847, 2004.

\bibitem{Pavlidi2013}
D.~Pavlidi, A.~Griffin, M.~Puigt, and A.~Mouchtaris, ``Real-time multiple sound
  source localization and counting using a circular microphone array,''
  \emph{IEEE Transactions on Audio, Speech, and Language Processing}, vol.~21,
  no.~10, pp. 2193--2206, 2013.

\bibitem{Loesch2008}
B.~Loesch and B.~Yang, ``Source number estimation and clustering for
  underdetermined blind source separation,'' in \emph{International Workshop on
  Acoustic Signal Enhancement (IWAENC), Seattle, Washington, USA}, 2008.

\bibitem{Catalbas2017}
M.~C. Catalbas and S.~Dobrisek, ``{3D} moving sound source localization via
  conventional microphones,'' \emph{Elektronika ir Elektrotechnika}, vol.~23,
  no.~4, pp. 63--69, 2017.

\bibitem{Traa2013}
J.~Traa and P.~Smaragdis, ``Blind multi-channel source separation by
  circular-linear statistical modeling of phase differences,'' in \emph{IEEE
  International Conference on Acoustics, Speech and Signal Processing
  (ICASSP)}.\hskip 1em plus 0.5em minus 0.4em\relax IEEE, 2013, pp. 4320--4324.

\bibitem{Gala2019Moving}
D.~Gala and L.~Sun, ``Moving sound source localization and tracking using a
  self rotating bi-microphone array,'' in \emph{ASME 2019 Dynamic Systems and
  Control Conference}.\hskip 1em plus 0.5em minus 0.4em\relax American Society
  of Mechanical Engineers Digital Collection, 2019.

\bibitem{Gala2019}
D.~Gala, N.~Lindsay, and L.~Sun, ``Realtime active sound source localization
  for unmanned ground robots using a self-rotational bi-microphone array,''
  \emph{Journal of Intelligent \& Robotic Systems}, vol.~95, no. 3-4, pp.
  935--954, 2019.

\bibitem{Ester1996}
M.~Ester, H.-P. Kriegel, J.~Sander, and X.~Xu, ``A density-based algorithm for
  discovering clusters in large spatial databases with noise,'' in \emph{Kdd},
  vol.~96, no.~34, 1996, pp. 226--231.

\bibitem{Knapp1976The}
C.~Knapp and G.~Carter, ``The generalized correlation method for estimation of
  time delay,'' \emph{IEEE Transactions on Acoustics, Speech, and Signal
  Processing}, vol.~24, no.~4, pp. 320--327, Aug 1976.

\bibitem{azaria1984time}
M.~Azaria and D.~Hertz, ``Time delay estimation by generalized cross
  correlation methods,'' \emph{IEEE Transactions on Acoustics, Speech, and
  Signal Processing}, vol.~32, no.~2, pp. 280--285, 1984.

\bibitem{naylor2010speech}
P.~Naylor and N.~D. Gaubitch, \emph{Speech dereverberation}.\hskip 1em plus
  0.5em minus 0.4em\relax Springer Science \& Business Media, 2010.

\bibitem{Gala2010}
D.~R. Gala, A.~Vasoya, and V.~M. Misra, ``Speech enhancement combining spectral
  subtraction and beamforming techniques for microphone array,'' in
  \emph{Proceedings of the International Conference and Workshop on Emerging
  Trends in Technology (ICWET)}, 2010, pp. 163--166.

\bibitem{Gala2011}
D.~R. Gala and V.~M. Misra, ``{SNR} improvement with speech enhancement
  techniques,'' in \emph{Proceedings of the International Conference and
  Workshop on Emerging Trends in Technology (ICWET)}.\hskip 1em plus 0.5em
  minus 0.4em\relax ACM, 2011, pp. 163--166.

\bibitem{ISO12001}
``{International Organization for Standardization ({ISO}), British, European
  and International Standards ({BSEN}), Noise emitted by machinery and
  equipment -- Rules for the drafting and presentation of a noise test code},''
  \emph{12001: 1997 Acoustics}.

\bibitem{Hansen2001}
B.~Goelzer, C.~H. Hansen, and G.~Sehrndt, \emph{Occupational exposure to noise:
  evaluation, prevention and control}.\hskip 1em plus 0.5em minus 0.4em\relax
  World Health Organisation, 2001.

\bibitem{Calmes2009}
L.~Calmes, ``Biologically inspired binaural sound source localization and
  tracking for mobile robots.'' Ph.D. dissertation, RWTH Aachen University,
  2009.

\bibitem{Raj2017}
C.~D. Raj, ``Comparison of {K} means {K} medoids {DBSCAN} algorithms using
  {DNA} microarray dataset,'' \emph{International Journal of Computational and
  Applied Mathematics (IJCAM)}, 2017.

\bibitem{Farmani2017}
N.~Farmani, L.~Sun, and D.~J. Pack, ``A scalable multitarget tracking system
  for cooperative unmanned aerial vehicles,'' \emph{IEEE Transactions on
  Aerospace and Electronic Systems}, vol.~53, no.~4, pp. 1947--1961, Aug 2017.

\bibitem{Celebi2013}
M.~E. Celebi, H.~A. Kingravi, and P.~A. Vela, ``A comparative study of
  efficient initialization methods for the k-means clustering algorithm,''
  \emph{Expert systems with applications}, vol.~40, no.~1, pp. 200--210, 2013.

\bibitem{Fischler1981}
M.~A. Fischler and R.~C. Bolles, ``Random sample consensus: a paradigm for
  model fitting with applications to image analysis and automated
  cartography,'' \emph{Communications of the ACM}, vol.~24, no.~6, pp.
  381--395, 1981.

\bibitem{Donohue2009}
K.~D. Donohue, ``Audio array toolbox,'' \emph{[Online] Available:
  \url{http://vis.uky.edu/distributed-audio-lab/about/} , 2019, May 20}.

\bibitem{allen1979image}
J.~B. Allen and D.~A. Berkley, ``Image method for efficiently simulating
  small-room acoustics,'' \emph{The Journal of the Acoustical Society of
  America}, vol.~65, no.~4, pp. 943--950, 1979.

\bibitem{Corpus}
K.~D. Donohue, ``Audio systems lab experimental data - single-track
  single-speaker speech,'' \emph{[Online] Available:
  \url{http://web.engr.uky.edu/~donohue/audio/Data/audioexpdata.htm} , 2019,
  May 20}.

\end{thebibliography}

\end{document}